\documentclass[graybox]{svmult}

\usepackage{mathptmx}       
\usepackage{helvet}         
\usepackage{courier}        
\usepackage{amsmath,bm}
\usepackage{amssymb}
\usepackage{subfigure}
\usepackage{blindtext}
\usepackage{hyperref}
\usepackage{amsfonts}
\usepackage{booktabs}
\usepackage{array}
\usepackage[dvipsnames]{xcolor}
\usepackage{tikz}
\usepackage{float}
\usepackage{type1cm}        
\usepackage{natbib}
\usepackage{makeidx}         
\usepackage{graphicx}        
\usepackage{multicol}        
\usepackage[bottom]{footmisc}
\usepackage{colortbl,booktabs}
\usepackage{tablefootnote}

\newcolumntype{M}[1]{>{\centering\arraybackslash}m{#1}}

\makeindex             

\begin{document}
\title*{Mixture polarization in inter-rater agreement analysis: a Bayesian nonparametric index}
\titlerunning{Mixture polarization in inter-rater agreement analysis} 
\author{Giuseppe Mignemi, Antonio Calcagnì, Andrea Spoto, Ioanna Manolopoulou}
\authorrunning{Mignemi et al.} 
\institute{Giuseppe Mignemi \at University of Padova, 
\email{giuseppe.mignemi@phd.unipd.it}\\
\url{https://github.com/MignemiG}
\and Ioanna Manolopoulou \at University College London,  
\email{i.manolopoulou@ucl.ac.uk}
\and Antonio Calcagn\`i, \at DPSS, University of Padova,
\email{antonio.calcagni@unipd.it}\\
GNCS Research Group, National Institute of Advanced Mathematics (INdAM)
\and Andrea Spoto \at University of Padova,  
\email{andrea.spoto@unipd.it}
}
%
%
\maketitle

\abstract{In several observational contexts where different raters evaluate a set of items, it is common to assume that all raters draw their scores from the same underlying distribution. However, a plenty of scientific works have evidenced the relevance of individual variability in different type of rating tasks. To address this issue the intra-class correlation coefficient (ICC) has been used as a measure of variability among raters within the Hierarchical Linear Models approach. A common distributional assumption in this setting is to specify hierarchical effects as independent and identically distributed from a normal with the mean parameter fixed to zero and unknown variance. The present work aims to overcome this strong assumption in the inter-rater agreement estimation by placing a Dirichlet Process Mixture over the hierarchical effects' prior distribution. A new nonparametric index $\lambda$ is proposed to quantify raters polarization in presence of group heterogeneity. The model is applied on a set of simulated experiments and real world data. Possible future directions are discussed. 
}

%
%
\section{Introduction}
\label{sec:1}
 In several contexts, decision-making relies heavily (or exclusively) on expert ratings, especially in situations where a direct quantification of quality of an object or a subject is either  impossible or unavailable. Examples include applicant selection procedures, grading of student assignments in education, or risk evaluation in emergencies, all of which rely on  observational ratings made by experts. For ease of exposition, throughout this paper we will refer to evaluation of students' work in an educational context as the primary example. To ensure consistency across different teachers, harmonization of marking criteria is often used to improve inter-rater agreement and homogeneity \citep{Gisev,Gwet_Li}; however, discrepancies between grades assigned by different teachers may still persist \citep{Bygren_2020, Barneron_2019, Makransky_2019, Zupanc2018}, reflecting each teacher's approach to evaluation.  Therefore, statistical models that can capture inter-rater agreement (or disagreement) can shed light on heterogeneity between teachers and aid the mark moderation process \citep{Bygren_2020,Barneron_2019,Crimmins_2016}. \\
The specific context that we are considering in this work is the observational setting where a set of raters are evaluating different sets of items \footnote{Commonly referred to as \textit{subjects} in the rating context.} out of a total population of items; these 
sets may be completely disjoint (i.e., each item is evaluated by exactly one rater). Each item is represented by a set of covariates, assumed to follow some distribution. Within a hierarchical statistical model, a common assumption is that raters (who may or may not include covariates) 
may each be characterized through a latent variable capturing e.g. whether an evaluator is generous or how they assess different aspects of the work. In the simplest setting, in an evaluation context where there is no space for subjectivity, these latent variables will be identical for all raters, in the sense that their view of the item is identical and as a result their evaluation style is assumed to be the same. However, it is well-known in many scientific fields, e.g., cognitive neuroscience \citep{Barneron_2019,Makransky_2019,Briesch_2014}, statistics \citep{Agresti2015,Gelman2013} and psychometrics \citep{Martinkova,Nelson_2015,Hsiao_2011}, that individual variability in rating tasks \citep{Wirtz2020} needs to be accounted for when aggregating or interpreting individual raters' recommendations. \\
Existing works account for heterogeneity between raters through a latent variable within a mixed-effects model \citep{Patrícia_2023,Martinkova, Nelson_2015, Nelson2008}. In other words, a regression model is used where the rating is modelled conditionally on covariates with a random effect that varies across raters. However, the distribution of the latent variable is typically assumed to be unimodal, and cannot capture eg. polarisation or clustering of rater types. The present work aims to extend these models to account for clustered variability between raters. Through a Bayesian approach, a Dirichlet process mixture prior is placed over the hierarchical rater effects in a linear model. This flexible prior naturally accommodates different clusters among raters (i.e., different distributions for the rater effects). A multiple-level model is specified in which observations (i.e., ratings) are nested within raters, and in turn these are nested within clusters. These clusters reflect distinct groups of raters in terms of their decision-making, and can be used to characterise  the level of (dis)agreement. 
The level of multimodality (i.e., how separated the latent group densities are) quantifies the polarization of the latent groups. For instance, a large variance between teacher scores might be due to both the presence of two main divergent latent trends among them or to a high level of noise in their assessing \citep{Koudenburg_2022}. It is important to differentiate the two cases and quantify the group polarization both for theoretical and practical purposes. Differentiate systematic differences of opinion against high level of noise might be needed \citep{Koudenburg_2021} . They are two very different cases and much attention must be paid in distinguishing one another. The former is a case of high group polarization \citep{Esteban_1994}: two different teachers clusters emerge with a small within-cluster variance and a large variance between different clusters. In the second case only one cluster emerge with a large variance. 
 It might be argue that in the first case, even that the overall agreement might be quite low since there are two main different trends among raters, there might be a high agreement within the same trend \citep{Tang2022}. Assuming the latent agreement among raters as the degree of latent similarity in rating, an index regarding the polarization of the different possible groups of raters might be informative \citep{Koudenburg_2022, Tang2022, Koudenburg_2021}.  
In this work we introduce a novel index to quantify the latent polarization among raters through the posterior distribution of the hierarchical effects \citep{DiMaggio_1996}. It naturally derives from the nonparametric model and overcomes some strong assumptions (e.g., the number of latent groups, the ratings distribution) of the previous indices \citep{Koudenburg_2021,Esteban_1994}. This nonparametric index, referred to as $\lambda$ index, is based on the shape of the posterior distribution of the hierarchical effects. It connects two different research lines: it relates the works on distribution polarization of opinions \citep{Koudenburg_2022, Tang2022, Koudenburg_2021} with those about the inter-rater agreement analysis \citep{Patrícia_2023,Martinkova,Nelson_2015,Gisev,Gwet_Li,Nelson2008}. \\
The paper proceeds as follows: Section ~\ref{sec:2} is devoted to the general psychometric framework, the key concepts of inter-rater agreement, inter-rater reliability are introduced; the statistical model is specified in Section ~\ref{sec:3} and the adopted Gibbs sampler in Section \ref{sec:4}; the novel rater similarity  index is described in Section ~\ref{sec:5}; simulation studies are reported in Section ~\ref{sec:6}, as an illustrative example, a real data analysis is described in Section ~\ref{sec:7}; it is followed by conclusion and future directions in Section \ref{sec:8}.    

\section{Existing work in inter-rater agreement and hierarchical effects models}
\label{sec:2}
Several methods and statistical models that aim to account for inter-rater variability have appeared in the literature  \citep{Nelson_2015,Gwet_Li,cicchetti_1976}. 
Models such as the Cultural Consensus Theory \citep{CCT}, which explores individuals’ shared cultural knowledge, have been proposed to capture unobserved agreement and similar trends in groups of raters \citep{CCT2}. 
Two related but different concepts have been introduced: \textit{inter-rater agreement} and \textit{inter-rater reliability}. The former refers to the extent to which different raters' evaluations are concordant (i.e, they assign the same value to the same item), whereas the latter refers to the extent to which their evaluations consistently distinguish different items \citep{Gisev}. In other words, while the inter-rater agreement indices quantify the \textit{observed} concordance, the inter-rater reliability indices aim to quantify the \textit{consistency} of their evaluations (e.g., despite assigning different values, the distinction among the items is the same). The present work focuses on latent agreement intended as homogeneity in the evaluators' point of view \citep{Tang2022, Esteban_1994}. 
\\
A number of methods are available to quantify both inter-rater agreement and inter-rater reliability. Indices for pairs \citep{Nelson2008,McHugh12} or multiple raters \citep{Jang}, for binary \citep{Gwet_Li}, polytomous \citep{Nelson_2015} or continuous \citep{Liljequist} ratings are commonly used in different contexts. Recent developments using the framework of Hierarchical Linear Models (i.e., HLMs) provide a more accurate estimation of inter-rater reliability accounting for different sources of variability \citep{Patrícia_2023}. \\
Despite the popularity of work on this issue, less attention has been paid to possible latent similarities of the raters \citep{Wirtz2020}. From a psychometric point of view, it can be appealing to assess the extent to which different raters might be heterogeneous in their ratings \citep{Patrícia_2023,Martinkova,Koudenburg_2021,Casabianca_2015,Nelson_2015,Gisev,DeCarlo_2008,Gwet_Li,Nelson2008}. 

There are certain situations in which the subjective opinion of the raters is very informative; as a simple example, the type of teachers' training or experience can be thought of as latent states which affect a range of evaluations differently \citep{Tasha_2023,Barneron_2019,Bonefeld_2018, Dee_2005}. Sometimes the major interest is not on the mere consistency between raters, but on their actual evaluation. For instance, in a selection process the actual students' scores are very relevant for their admission \citep{Zupanc2018}. Even if a strict standardization of teachers evaluation is not feasible, some statistical methods can tackle these issues.   
In all these contexts the assessment of uniformity among raters could be useful and would provide further information about the rating process.\\ 
To this aim, existing work, e.g.\cite{Patrícia_2023,Nelson_2015,Casabianca_2015,Hsiao_2011, Cao_2010, DeCarlo_2008},  adopts an hierarchical approach where correlations between ratings are naturally captured through an hierarchical Bayesian model. 
Each rater $i = 1,..,I$ is assumed to be rating a different set of items $\mathcal{J}_i \in \mathcal{J}$, $\mathcal{J}_i \cap \mathcal{J}_{i+1} = \emptyset$\footnote{The multiple rating case (i.e., raters rate the same set of items, $\mathcal{J}_i = \mathcal{J}$, $i=1,\dots,I$) is addressed in Appendix.}. 
The rating $y_{ij}$ of the item $j \in \mathcal{J}_i$ carried out by rater $i = 1,..,I$, is modelled as follows:
\begin{eqnarray}\label{HLM}
    y_{ij}&=&\mathbf{x}_{ij}'\mathbf{\beta}+\mathbf{z}_{ij}'\mathbf{u}_i+\epsilon_{ij},\;\; i = 1,..,I, \; j \in \mathcal{J}_i. \;\; 
    \end{eqnarray}\\
Here $\mathbf{x}_{ij}$ and $\mathbf{z}_{ij}$ are, respectively, $1 \times p$ and  $1 \times q$ vectors of distinct explanatory variables of rating $y_{ij}$; $\mathbf{\beta}$ is a $p \times 1$ vector of non varying  effects and $\mathbf{u}_i$ is a $q \times 1$ vector the hierarchical effects of rater $i$. 
 \\
In the standard HLM formulation, the following distribution is specified for the rater effects: 
 \begin{eqnarray}   
    \mathbf{u}_i &\sim& N_q(\mathbf{0}, \pmb{\Sigma}), \quad  i = 1,..,I. \nonumber
\end{eqnarray}
Where $N_q(\cdot)$ stands for a $q$-variate normal distribution; Here $\mathbf{0}$ is a $q \times 1$ zero vector and $\pmb{\Sigma}$ is a  $q \times q$ positive semi-definite covariance matrix. For the hierarchical normal linear model $\epsilon_{ij} \sim N(0,\sigma_\epsilon)$, with $\mathbf{u}_i$ and $\epsilon_{ij}$ typically assumed independent. 
The distribution of each vector-valued hierarchical effects $\mathbf{u}_i$ is then assumed to follow some distribution and captures variability across different raters.\\
In the above mentioned example, $y_{ij}$ is the score given to student $j \in \mathcal{J}_i$'s essay by teacher $i$. Since an observational approach is adopted (i.e., each raters rates a different set of items), the effect of the student is not identifiable (each student is rated only by one rater). Assuming that students effects are i.i.d., their variance is added to that of the residuals.  
In the univariate case (i.e., when $z_i=1$, varying intercept model) the relevance of the raters effect $u_i \sim N(0,\sigma^2_u)$, where $\sigma^2_u>0$ is the variance, might be quantified through the \textit{intraclass correlation coefficient} (i.e., ICC):  
 \begin{center}
    $ICC= \frac{\sigma_u^2}{\sigma_u^2+\sigma_\epsilon^2}$.
\end{center}
\vspace{5px}
It is the ratio between the variance of the raters effect and the total variability of the model, i.e.,  the proportion of variance of the score due to the teacher, which reflects the correlation of two ratings given by the same rater. Smaller values of ICC indicate a small effect of the rater on the student's score. 

\section{Dirichlet Process Mixture and hierarchical effects}
\label{sec:3}
The HLM  assumption regarding the distribution of the hierarchical effects
is crucial in characterising different possible clusters or latent patterns of heterogeneity among raters \citep{Dorazio_2009}. The common Gaussian assumption for the distribution of the these effects  may obscure skewness and multimodality present in the data.
A more flexible specification of the hierarchical effects distribution can help capture more complex patterns of variability. Models that account for skew-normal  \citep{Lin_2008}, skew-normal-cauchy \citep{Kahrari_2019}, multivariate \textit{t} \citep{Wang_2014}, extreme values \citep{McCulloch_2021} effects distributions have been proposed \citep{Schielzeth_2020}. Nevertheless, they poorly account for the possible presence of multimodality in those distributions. In this regard, a mixture distribution has been proposed as a potential solution \citep{Heinzl_2013,Kyung_2011,Kim_2006}. Each mode can then correspond to a cluster with a similar pattern (e.g., the same deviation from the population mean). Several works have explored this issue in the past two decades \citep{Villarroel_2009, Tutz_2017}. 
For instance, Verbeke and Lesaffre \citep{Verbeke_1996} proposed a standard normal mixture distributions for the hierarchical effects. James and Sugar \citep{James_2003} explored this approach in the context of functional data. De la Cruz-Mes{\'i}a \citep{DelaCruz_2006} proposed a mixture distribution for non-linear hierarchical  effects in modelling continuous time autoregressive errors. A heteroscedastic normal mixture model in the hierarchical effects distribution was considered in linear \citep{Komarek_2010} and generalized hierarchical linear \citep{Komarek_2012} models.   
Despite the breadth of specifications for the mixture model, in all the aforementioned models, the number of mixture components needs to be specified. Although this may not be a critical assumption in certain contexts, it may be questionable or detrimental in settings with a lack of a priori information on the level of multimodality, especially in cases where the characterisation of the multimodality is of direct interest.\\
When the number of components of the mixture is unknown, a Dirichlet Process Mixture (hereafter DPM) for the hierarchical effects is a natural extension \citep{Gill_2009,Navarro_2006, Verbeke_1996}. This nonparametric extension allows the model to capture an unknown marginal distribution of the hierarchical effects through the Dirichlet Process  \citep{Antoniak_1974,Ferguson_1973}. 
Modeling the hierarchical effect $\mathbf{u}_i$ as an infinite mixture of some distribution family (e.g., Normal) enables the model to account for possible multimodality without specifying the number of mixture components. Some existing works adopted this nonparametric approach and pose a DPM prior over the hierarchical effects (e.g., \cite{Heinzl_2013, Heinzl_2012, Kyung_2011}). \\
The HLM of Equation (\ref{HLM}) is then specified in the same way as before through: \\
\begin{eqnarray}
    y_{ij}&=&\mathbf{x}_{ij}'\mathbf{\beta}+\mathbf{z}_{i}'\mathbf{u}_i+\epsilon_{ij},  \quad \quad  i = 1,..,I, \quad j \in \mathcal{J}_i. \;\; \nonumber
\end{eqnarray}
The following hierarchical prior distribution is placed over the raters effects:
 \begin{eqnarray}   
   \mathbf{u}_i | \mathbf{\mu}_{i}, \mathbf{Q}_{i} &\sim& N_q(\mathbf{\mu}_{i}, \mathbf{Q}_{i}) \nonumber\\ 
    \mathbf{\mu}_i,\mathbf{Q}_i | G &\stackrel{iid}{\sim} & G \nonumber\\
    G &\sim& DP(\alpha, G_0)\nonumber
\end{eqnarray}
where $\mathbf{\mu}_i$ and $ \mathbf{Q}_i$ are, respectively, the $q \times 1$ a location parameter vector and the $q \times q$ positive semi-definite covariance matrix for the hierarchical effects $\mathbf{u}_i$ of rater $i=1,\dots,I$. Here $\epsilon_{ij} \sim N(0,\sigma_\epsilon)$, $i = 1,..,I$, $j \in \mathcal{J}_i$; $\mathbf{u}_i$ and $\epsilon_{ij}$ are assumed independent as before. \\

\subsection{DPM as a generative process for the hierarchical effects}
Here, $DP(\alpha, G_0)$ is a DPM with  $\alpha>0$ \textit{precision parameter} and \textit{base measure} $G_0$. These specify the mixing distribution $G$ \citep{Heinzl_2013}, so that  each realization of $G$ is almost surely a discrete 
 probability measure on the space $(\Omega, \mathcal{F})$ \citep{Blackwell_1973}. Thus, since the DPM is a discrete generative process with non-zero probability of ties, some of the realizations might be identical to each other with probability determined by the precision parameter $\alpha$. Therefore, specifying this hierarchical model on the components location parameters $\mathbf{\mu}$ induces a clustering in the hierarchical effects (i.e., the raters) \citep{Kyung_2011}; hierarchical effects belonging to the same $c$-th cluster  with location parameter $\mathbf{\mu}_c$ are then independent and identically distributed. 
 In other words, in the context of the HLM, the DPM specifies the component-specific location parameter $\mathbf{\mu}_c$, so that each rater has each has their own unique hierarchical effects value $\mathbf{u}_i$ \citep{Heinzl_2013}.\\ 
The DPM is a generative process commonly used in conjunction with a parametric family of distributions (e.g., Normal, Poisson), and  the base measure parameter $G_0$ denotes this specified distribution. Thus, for any element $A_n$, $n=1,..,N$, of $\mathcal{A}$, a finite measurable partition of $\Omega$, 
\begin{center}
    $(G(A_1), G(A_2),...,G(A_N) ) \sim Dir(\alpha G_0(A_1),\alpha G_0(A_2),...,\alpha G_0(A_N))$ 
\end{center}
\vspace{1.5px}
where $Dir(\cdot)$ stands for the Dirichlet distribution, and
$G_0$ defines the expectation of $G$\footnote{
Considering the partition $(A,A^c)$ of $\Omega$ and thus that $G(A) \sim Be(\alpha G_0(A),\alpha G_0(A^c))$ the expectation of $G(A)$ is defined as: 
 $\mathbb{E}[G(A)]=\frac{\alpha G_0(A)}{\alpha G_0(A)+\alpha G_0(A^c)}=
 \frac{\alpha G_0(A)}{\alpha (G_0(A)+G_0(A^c))}=G_0(A). 
 $

}, therefore they have the same support. The parameter $\alpha$, a multiplicative constant of the vector-valued Dirichlet parameter, determines the probability of a new realization of the process to be different of the previous ones \citep{Blackwell_MacQueen}. In other words, it governs the probability that the DPM generates a new cluster. Formally, the generative property of the DPM is that, for $i=1,...,I$, with $I$ being for instance the total number of raters:
\begin{center}
    $G \sim DP(\alpha, G_0)$, \quad $\mathbf{\mu}_i|G \sim G$
\end{center}
\vspace{1.5px}
the probability that the new $I$-th realization $\mu_I$ of $G$ assumes a different values than the previous ones is described by the well known \textit{P\'{o}lya Urn Model}:
\begin{center}
    $\mathbf{\mu}_I|\mathbf{\mu}_1,\mathbf{\mu}_2,...,\mathbf{\mu}_{I-1},\alpha \sim \frac{\alpha}{\alpha+I-1}G_0+\frac{1}{\alpha+I-1}
    \displaystyle\sum_{c=1}^{C} r_c$
\end{center}
\vspace{1.5px}
with $C \in \mathbb{N}$ being the number of already observed distinct clusters among the realizations of $G$ (i.e., the number of the different values of $\mathbf{\mu}$ already observed, in other words the number of clusters) and $r_c$ counts the elements in the $c$-th cluster. Basically, since $G$ is a discrete probability measure, the $C$ clusters represent different point masses (or different sets of point masses in the multivariate case) and $r_c$ is the frequency of each of them.  Considering the conditional distribution of $\mu_I$ as a mixture distribution, the probability that $\mathbf{\mu}_I$ is a new point mass sampled from $G_0$ is proportional to $\alpha$, the probability that it is equal to the already observed $c$-th point mass is proportional to $r_c$. In this notation, the role of $\alpha$ in sampling a new (not already observed) value of $\mathbf{\mu}_I$ (i.e., a new point mass, a new cluster) is interpretable.\\
To this regard, \cite{Sethuraman} described a \textit{stick-breaking construction} of the DP \footnote{Other stick-breaking representations might be used, e.g., \cite{Rigon2021,Stefanucci2021,Rodriguez_2011}.}. In this formulation G is equivalent to:
\begin{center}
    $G=\displaystyle\sum_{c=1}^{\infty} \pi_c \delta_{\mathbf{\mu}_c}$
\end{center}
\vspace{1.5px}
where $\delta$ is the Dirac measure on $\mathbf{\mu}_c $ and $\mathbf{\mu}_c \stackrel{iid}{\sim} G_0$ is assumed. The weights $\{\pi_c\}_{c=1}^{\infty}$ of the infinite mixture result from the stick-breaking procedure as follows:
\begin{center}
    $\pi_c= \nu_c \prod_{l<c}(1-v_l)$
 \end{center}
 \begin{center}
    $v_c  \stackrel{iid}{\sim} Be(1,\alpha)$
\end{center}
\vspace{1.5px}
with $Be(\cdot)$ indicating the Beta distribution and $\{\nu_c\}_{c=1}^{\infty}$ being reparameterized weights. It is even more explicit in this construction that the random measure $G$ is a mixture of point masses. The distribution of the random weights $\mathbf{\pi}$ (i.e., the probability of different allocation to the clusters) is governed through the stick-breaking process by the precision parameter $\alpha$. Further details are given in the Appendix.\\
In practice, one of the established approximations to the stick-breaking process is to truncate the infinite number of components to a large, finite value: 
\begin{center}
    $G=\displaystyle\sum_{c=1}^{R} \pi_c \delta_{\mathbf{\mu}_c}$
\end{center}
\vspace{1.5px}
for large enough value of $R$ \citep{Tutz_2017,Gelman2013}.\\
In summary, the hierarchical effects distribution considering a stick breaking construction of the DPM might be then specified as follow: 
\begin{eqnarray}
\mathbf{u}_i | \mathbf{\mu}, \mathbf{Q}, &\stackrel{iid}{\sim} &\displaystyle\sum_{c=1}^{R} \pi_c N_q(\mathbf{\mu}_c, \mathbf{Q}_c), \;\;\;i = 1,\ldots, I \nonumber \\
\mathbf{\mu}_c, \mathbf{Q}_c &\stackrel{iid}{\sim}& G_0 \nonumber \\
\pi_c&=& \nu_c \prod_{l<c}(1-v_l),\;\;\textrm{where} \nonumber \\
v_c  &\stackrel{iid}{\sim}& Be(1,\alpha), \;\;c=1\ldots,R \nonumber
\end{eqnarray}
With this nonparametric model specification, latent common tendencies among raters might emerge through the components of the model \citep{Heinzl_2013, Heinzl_2012, Kyung_2011}. The Bayesian approach allows us to characterize the shape of the distribution of the rater effects, as well as explore the effect of uncertainty on these \citep{Gelman2013}. For example, in an applied context, strict vs. accommodating are very common latent states that drive students' essays grading process \citep{Zupanc2018,Briesch_2014, Dee_2005}.


\section{Prior distributions and estimation procedure}
\label{sec:4}
The DPM mixture model has been well studied in the literature in a variety of different settings, especially within Bayesian inference \citep{Canale_2017,Muller_2015}. Several sampling schemes have been proposed both in the Bayesian context (e.g., \cite{Canale_2011,Dahlin_2016,Kyung_2011}) and in the frequentist one (e.g., \cite{Tutz_2017}). 
Within the Bayesian framework, Gibbs sampling \citep{Dahlin_2016}, slice sampler \citep{Kyung_2011,Walker_2007}, Sequential Monte Carlo algorithms \citep{Ulker_2010}, split-merge algorithms \citep{Bouchard-Côté_2017}, have been proposed among others. \\
In this work, the model specification permits the use of conjugate priors, so that a blocked Gibbs sampling can be used \citep{Heinzl_2013, Heinzl_2012, Kyung_2011}., with details shown below. \\

\subsection{Prior specification}
\label{sub:1}
Several of the parameters in the model have conjugate prior distributions which allow easier computation. 
\begin{itemize}
    \item For the effects $\mathbf{\beta}$ the following hierarchical prior is assigned: \\
\begin{eqnarray}
   \mathbf{\beta}|\mathbf{b}_{\beta}, \mathbf{B}_{\beta} &\sim & N_p(\mathbf{b}_{\beta}, \mathbf{B}_{\beta})\nonumber\\
    \mathbf{b}_{\beta} &\sim& N_p(\mathbf{b}_0, \mathbf{S}_0)\nonumber\\
    \mathbf{B}_{\beta}&=&diag(\sigma_{\beta_1}^2,...,\sigma_{\beta_p}^2)\nonumber\\
    \sigma_{\beta_m}^2 &\stackrel{iid}{\sim}& IG(a_{\beta_0}, b_{\beta_0} )\nonumber
\end{eqnarray}
for $m =1,...,p$, where $p$ is the number of covariates associated to the effects $\mathbf{\beta}$. Here, $IG(\cdot)$ stands for inverse-gamma with shape parameters $a_{\beta_0}>0$ and rate parameters $b_{\beta_0}>0$. 
Where $\mathbf{b}_0$ and $\mathbf{S}_0$ are, respectively, the $p \times 1$ vector of location parameters and the  $p \times p$ positive semi-definite covariance matrix of $\mathbf{b}_\beta$ (i.e, the location parameter vector of the non varying effect $\mathbf{\beta}$); $\mathbf{b}_\beta$ and $\mathbf{S}_\beta$ are, respectively, the $p \times 1$ location parameter and the $p \times p$ positive semi-definite covariance matrix for $\mathbf{\beta}$ (i.e., the non varying effect). 
The set $\{\mathbf{b}_0, \mathbf{S}_0,a_{\beta_0}, b_{\beta_0}\}$ of the hyperparameters are specified by the user. 
A diagonal matrix is suggested for $\mathbf{S}_0$ as showed by \cite{Heinzl_2012}.\\
\item 
A diagonal structure for the $q \times q$ prior covariance matrix $\mathbf{Q}_r$ for the hierarchical effects is specified as follows for each mixture component $r = 1,...,R$ and each each related covariate $d =1,..., q$: 
\begin{eqnarray}
   \mathbf{Q}_r &=& diag(\sigma_{Q_{1r}}^2,...,\sigma_{Q_{qr}}^2)\nonumber
\end{eqnarray}
For the base measure $G_0$\footnote{Assuming independence between the location and the scale parameters of each mixture component, and between all the scale parameters for each covariate $d=1,\dots,q$, $G_0$ is then the product of the $q$-variate normal and the $q$ inverse gamma distributions.} and the precision parameter $\alpha$ of the DP mixture model the following priors are specified: 
\begin{eqnarray}
   G_0&=&N_q(\mathbf{\mu}_0,\mathbf{D}_0) \times IG(a_{Q_0},b_{Q_0})^q
   \nonumber\\
   \mathbf{\mu}_0 &\sim& N_q(\mathbf{m}_0, \mathbf{W}_0)\nonumber\\
   \mathbf{D}_0 & = & diag(\sigma_{{D_0}_1}^2,...,\sigma_{{D_0}_q}^2)\nonumber\\
   \sigma_{{D_0}_d}^2 &\sim& IG(a_{D_0},b_{D_0})\nonumber \\
    \alpha &\sim& Ga(c_0^{\alpha},C_0^{\alpha})\nonumber 
\end{eqnarray}
\vspace{1.5px}
for $d=1,...,q$, where $q$ is the number of covariates associated to the hierarchical effects $\mathbf{u}_i$, and for $a_{D_0},a_{Q_0}>0$ and $b_{Q_0}, b_{D_0}>0$. Here $Ga(\cdot)$ stands for Gamma distribution with $c_0^{\alpha}>0$ and $C_0^{\alpha}>0$ respectively the shape and the rate parameters. 
Where $\mathbf{m}_0$ and $\mathbf{W}_0$ are, respectively, the $q \times 1$ location parameter vector and the $q \times q$ positive semi-definite covariance matrix of $\mathbf{\mu}_0$ (i.e. the location parameter of the base measure $G_0$); $\mathbf{\mu}_0$ and $\mathbf{D}_0$ are, respectively, the $q \times 1$ location parameter vector and the $q \times q$ positive semi-definite covariance matrix of the base measure $G_0$. 
The set $\{a_{Q_0},b_{Q_0}, \mathbf{m}_0, \mathbf{W}_0,a_{D_0}, b_{D_0}, c_0^{\alpha}, C_0^{\alpha}\}$ of the hyperparameters need to be fixed. A diagonal structure is suggested for $\mathbf{W}_0$ as above.\\
\item The following prior is assigned to the noise variance: 
\begin{eqnarray}
   \sigma_\epsilon &\sim& Ga(a_\epsilon,b_\epsilon)\nonumber
\end{eqnarray}
with $a_\epsilon >0$ and $b_\epsilon>0$ hyperparameters fixed by the user as well.
\vspace{1.5px}
\end{itemize}

\subsection{Posterior sampling}
Since most of the parameters in the model have conjugate prior distributions, a blocked Gibbs sampling algorithm was used for the posterior sampling \citep{Ishwaran2001} . \\
The parameter vector for the model is $\theta=\{
\mathbf{\beta}, \mathbf{b}_\beta,\mathbf{B}_\beta,
\mathbf{\mu}_0,\mathbf{D}_0,
\mathbf{Q}, 
\sigma_{\epsilon},
\alpha, \mathbf{\pi}, \mathbf{c} \}$ which is updated at each  state of the Markov chain of the Gibbs sampling. Here $\mathbf{c}= (c_1,\dots,c_I)$ is the allocation parameter of the raters to the clusters and. Further details on the following sampling are given in the Appendix. The closed-form marginal posteriors are as follows.\\
\begin{enumerate}
    \item Update parameters referring to effects $\mathbf{\beta}$: \\
       \begin{eqnarray}
  \mathbf{\beta}| \mathbf{b}_\beta,  \mathbf{B}_\beta, \mathbf{u}, \sigma_\epsilon , \mathbf{y} &\sim & N_p(\mathbf{b}_\beta^{*},\mathbf{B}_\beta^{*}) \nonumber
\end{eqnarray} 
For each covariate $m=1,...,p$ associated with a non varying  effect $\beta_m$, 
 \begin{eqnarray}
  b_{\beta_m}|\sigma_{\beta_m}^2, \beta_m &\sim & N \left( \left( \frac{1}{\sigma_{\beta_m}^2} + \frac{1}{s_{0_m}^2}\right)^{-1} \right( \frac{\beta_m}{\sigma_{\beta_m}^2} + \frac{m_{\beta_m}}{s_{0_m}^2} \left), \left( \frac{1}{\sigma_{\beta_m}^2} + \frac{1}{s_{0_m}^2}\right)^{-1}\right) \nonumber \\
  \sigma_{\beta_m}^2| b_{\beta_m}, \beta_m &\sim & IG\left(a_{b_0} + \frac{1}{2}, \frac{1}{2} (\beta_m-b_{\beta_m})\right) \nonumber
\end{eqnarray} 

    \item Update parameters referring to hierarchical effects: \\
\begin{itemize}
    \item For each rater $i = 1,...,I$:
 \begin{eqnarray}
  \mathbf{u}_i|\mathbf{\mu}_{c_i}, \mathbf{\mu}_0, \mathbf{Q}_0, \mathbf{\beta}, \sigma_\epsilon, \mathbf{y}_i &\sim & N_q(\mathbf{\mu}_{c_i}^*,\mathbf{Q}_{c_i}^*), \nonumber 
\end{eqnarray} 
\vspace{1.5px}
where $\mathbf{\mu}_{c_i}$ is the location parameter vector of the cluster where the $i$-th rater is allocated. \\[0.3cm]

\item For each component $r=1,...,R$ of the truncated mixture : \\[0.3cm]
- If $\nexists i : c_i = r$ (if no rater are currently allocated into cluster $r$), for each covariate $d =1,...,q$ associated to an hierarchical effect (independently):  
\begin{eqnarray}
\mathbf{\mu}_r | \mathbf{\mu}_0, \mathbf{D}_0 &\sim& N_q(\mathbf{\mu}_0, \mathbf{D}_0) \nonumber \\
  \sigma_{Q_{dr}}^2 &\sim& IG(a_{Q_0},b_{Q_0})\nonumber
\end{eqnarray} 
- If $\exists i : c_i = r$ (if at least one rater assigned to component $r$), for each covariate $d =1,...,q$ associated to an hierarchical effect (independently):
 \begin{eqnarray}
 \mu_{r_d} | \sigma_{Q_m}^2, \mu_{0_r}, \sigma_{D_{0_m}}^2, \mathbf{u}, \mathbf{c} &\sim& N(\mu_{0_r}^*, \sigma_{D_{0_m}}^{2*}) \nonumber \\
 \sigma_{Q_{dr}}^2|\mathbf{\mu}, \mathbf{u} &\sim& IG \left( a_{Q_0}^*, b_{Q_0}^*  \right) \nonumber
\end{eqnarray} 
Essentially, at each iteration $t$, if the $r$-th cluster is empty the component location parameters $\mu_r$ are sampled from the prior as suggested by \citep{Gelman2013}, otherwise they are drawn from the above mentioned closed-form posterior. 

\item Each rater $i =1,...,I$ is re-allocated into a cluster:
\begin{eqnarray}
 c_i| \mathbf{\pi}, \mathbf{\mu}, \mathbf{Q},\mathbf{u}_i &\sim& Cat(\mathbf{\omega}_i^*) \nonumber 
 \end{eqnarray} 
 where $Cat(\cdot)$ stands for Categorical distribution, and $\mathbf{\omega}_i^*$ is reported in the Appendix. 
\vspace{1.5px}
A truncated approximation for the DPM mixture model was used \citep{Gelman2013, Heinzl_2012} for a large value of $R$. The stick-breaking construction was used to generate the mixture weights $\pi_{1:R}$ .\\

\item For each component $r = 1,...,R-1$: 
 \begin{eqnarray}
 v_r|\mathbf{\phi}, \alpha &\sim& Be \left(1+c_r,\alpha+\displaystyle\sum_{l=c+1}^{R} r_l \right)
 \nonumber 
 \end{eqnarray}
 and $v_R=1$ for the last cluster. Here $c_r$ is the number of raters assigned to the cluster $r$, and $r_l$ is the number of raters assigned to the cluster $l$.\\
 \item The precision parameter is updated as follows: 
 \begin{eqnarray}
 \alpha|v_1,...,v_{R-1} &\sim& Ga\left(R-1+a_\alpha,b_\alpha-\displaystyle\sum_{c=1}^{R-1} ln(1-v_r) \right)
 \nonumber 
\end{eqnarray} 
\item For each covariate $d =1,...,q$ associated with an hierarchical effect the base measure parameters are updated: 
\begin{eqnarray}
 \mu_{0_d}|\sigma_{D_{0_d}}^2, \mathbf{\mu} &\sim& N\left( 
 \left( \frac{I}{\sigma_{D_{0_d}}^2} + \frac{1}{\sigma_{W_{0_d}}^2} \right)^{-1} 
 \left( \frac{I}{\sigma_{D_{0_d}}^2} \overline{\mu}_d + \frac{m_{0_d}}{\sigma_{W_{0_d}}^2} \right), \left( \frac{I}{\sigma_{D_{0_d}}^2} + \frac{1}{\sigma_{W_{0_d}}^2} \right)^{-1}
 \right) \nonumber 
\end{eqnarray}
where $\overline{\mu}_d$ is the mean of the location parameters related to the $d$-th covariate over all the clusters.
\begin{eqnarray}
 \sigma_{D_{0_m}}^2|\mu_{0_m},\mathbf{\mu} &\sim& IG \left( a_{D_0} + \frac{I}{2},  b_{D_0}+\frac{1}{2}\displaystyle\sum_{i=1}^{I} (\mu_{i_m}-\mu_{0_m})^2 \right) \nonumber
\end{eqnarray}
\end{itemize}
    \item Update the error variance: \\
\begin{eqnarray}
 \sigma_\epsilon^2|\mathbf{\beta}, \mathbf{u}, \mathbf{y} &\sim& IG \left( a_\epsilon + \frac{1}{2}I |\mathcal{J}|,  b_\epsilon+\frac{1}{2}\displaystyle\sum_{i=1}^{I} \displaystyle\sum_{j \in \mathcal{J}_i}\left(y_{ij}-\mathbf{X}_{ij} \beta-\mathbf{Z}_{ij} \mathbf{u}_i \right)^2 \right). \nonumber
\end{eqnarray}
Here $|\mathcal{J}|$ is the cardinality of the set of all the rated items $\mathcal{J}$, it equals the number of observations. 
\end{enumerate}
\vspace{1.5px}

\section{The nonparametric $\lambda$ index}
\label{sec:5}

\begin{figure}
    \centering
    \subfigure[]{\includegraphics[scale=0.27]{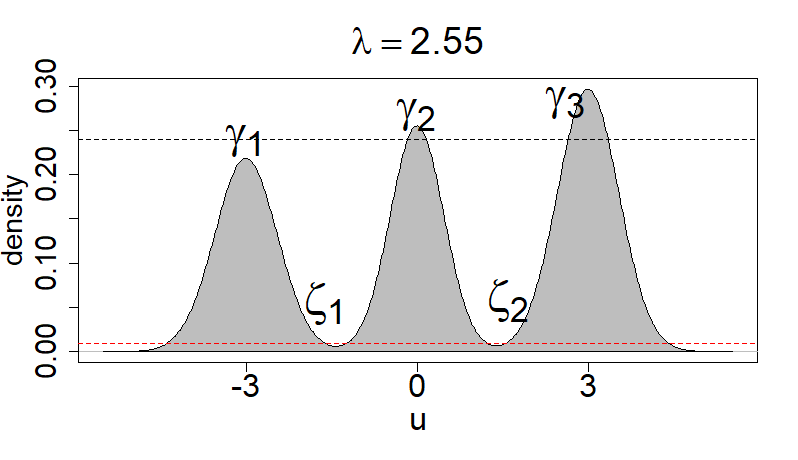}} 
    \subfigure[]{\includegraphics[scale=0.27]{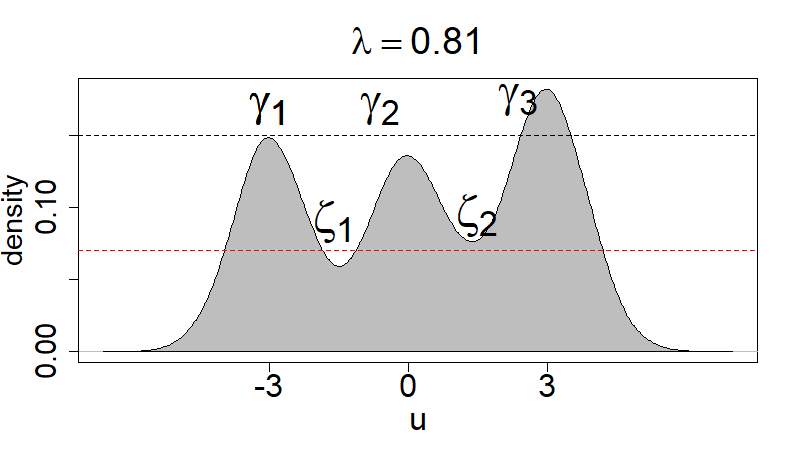}} 
    \subfigure[]{\includegraphics[scale=0.27]{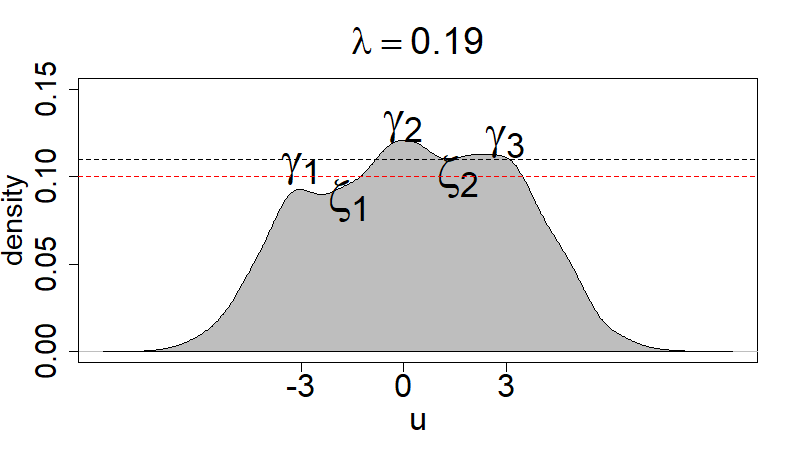}}
    \subfigure[]{\includegraphics[scale=0.27]{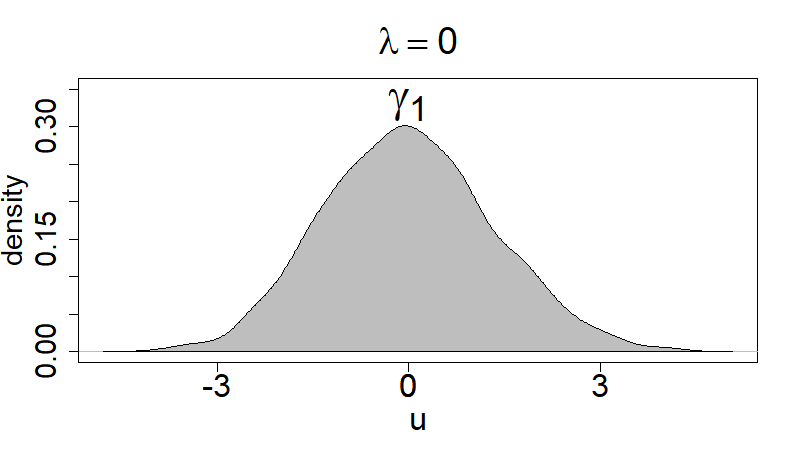}}
    \caption{Different values of $\lambda$ indicate different polarization levels. Three different values of $\lambda$ were computed for three different mixture distributions, respectively. The realizations of these distribution are here referred to as $u$. Black dotted lines indicate the mean mode density, red dotted lines indicate the mean antimode density. (a) High polarization: the mixture components are highly and clearly separate, the mean density values of the modes is far larger then the mean density value of the antimodes; the log-density ratio between these two quantities is $\lambda=2.55$ (b) Medium polarization: the mixture components are clearly separated, but the mean density values of the modes is closer to the mean density value of the antimodes; the log-density ratio between these two quantities is $\lambda=0.81$ (c) Low polarization: the mixture components are not clearly separated, the mean density values of the modes is very close to the mean density value of the antimodes; the log-density ratio between these two quantities is $\lambda=0.19$. (d) No polarization: the mixture distribution has only one mode (i.e., $\gamma_1$) and $\lambda=0$ since the number of mode is not greater then one.}
    \label{fig:1}
\end{figure}

The marginal posterior distribution of the hierarchical effects in the model outlined above captures information about the polarization or disagreement among raters (on the assumption that the model captures the data adequately). The ICC (i.e., intraclass correlation coefficient, \citep{Patrícia_2023,Martinkova, Agresti2015, Gelman2013}) might adequately quantify inter-rater variability if the normal distributional assumption of the rater hierarchical effect holds. 
Two assumptions are made computing the standard ICC considering a normal distributed hierarchical effect. Firstly, that the raters are sampled from the same population. Secondly, that possible different latent trends among raters are not interesting or eventually regarded as disagreement ratings. This might be a good first approximation of the rating process. Nevertheless, when more detailed considerations are needed, or subtle heterogeneity among raters is expected, the standard ICC might be less informative and inaccurate. Besides the latter issue, further information about the shape of the posterior might be quantified. For instance, in presence of a bimodal hierarchical effects distribution with two very distant modes (for example, when opinions are polarised), considering the posterior distribution of $\sigma_u$ as an index of variability among raters might be misleading. \\
Several indexes have been proposed to quantify group opinion polarization (e.g., \citep{Tang2022,Koudenburg_2022,Koudenburg_2021,Esteban_1994}) and to measure distribution bimodality (e.g., the Ashman’s D \citep{Forchheimer_2015} or the bimodal separation index \citep{Zang2003}). The strong assumptions behind their use limit them to be valid options only in the parametric context or when the number of clusters is known. A model based nonparametric index is here proposed to overcome these limitations. \\
To this end the full estimated distribution of $\mathbf{u}$ resulting from the model might be useful. At each iteration $t$, the density of $\mathbf{u}$ is given by the corresponding mixture model given the parameters at iteration $t$. Following the formulation of \citep{Gelman2013} , the set of modes and antimodes (i.e., the lowest frequent value between two modes) is identified. When the distribution of $\mathbf{u}$ is multimodal, the latent polarization (disagreement)  $\lambda$ is then defined as the log ratio between the mean density of the modes and the that of the anti-modes, it is zero when it is unimodal: \\
\begin{eqnarray}
    \lambda = \begin{cases}
        \log \left(\frac{\frac{1}{M}\displaystyle\sum_{m=1}^{M} f_{u}(\gamma_m)}{\frac{1}{M-1}\displaystyle\sum_{m=1}^{M-1} f_{u}(\zeta_m)}\right), & \textrm{if} \quad  M>1 \nonumber \\[1cm]
        0,  & \textrm{otherwise.} \nonumber \\
    \end{cases} 
\end{eqnarray}\\
Where $M$ is the number of modes $\gamma_m$, $m=1,\dots,M$ and the number of antimodes $\zeta_m$, $m=1,\dots,M-1$ of the density of $\mathbf{u}$; $f_{u}(\cdot)$ denotes the density at a specific point.  Larger values of $\lambda$ indicate strongly multimodal distribution of the hierarchical effects, whereas smaller values are evidence of weak multimodality, thus the estimated hierarchical effects are less concentrated.\\
As it is shown in Figure \ref{fig:1} larger values of $\lambda$ indicate distribution polarization, whereas smaller values indicate a less concentrated and more spread density distribution. The $\lambda$ index is strongly affected by both location and scale parameters of the mixture components. For this reason it might be very informative in presence of multimodal distributions. Assuming such a raters' group polarization as a result of low latent agreement among raters, the $\lambda$ index might be a useful diagnostic tool.   

\section{Simulation studies}
\label{sec:6}
The following simulations aim to evidence how the values of $\lambda$ varying across different polarization settings. The first simulation investigates the role of the precision parameter $\alpha$ and the variance of the mixture components in determining the values of $\lambda$. The second one shows the complementary role of $\lambda$ in the inter-rater agreement analysis and how this index varies across different settings.

\subsection{Simulation 1: DPM and $\lambda$}
\label{sub:1}
\subsubsection{Simulation setting}
The first simulation study explores the role that the precision parameter of the Dirichlet Process and the variance of the components have in determining the values of the log-density index $\lambda$. For simplicity purpose the mixture components are assumed to have the same variance $Q$ in this simulation, so the component subscription will be omitted. The objective is to study the effect of $\alpha$ and $Q$,  on  $\lambda$ conditional on all the other variables. 
Since the former has a crucial role in the determination on the point masses of $G$, and thus the concentration of its realizations, an inverse relation between $\alpha$ and $\lambda$ is expected if $Q$ is fixed. 
Likewise, an inverse relation between $Q$ and $\lambda$ is expected if $\alpha$ is fixed. It is interpretable as an index of the sharpness of the modes. For this reason both the precision parameter of the DPM and the variance of its components are expected to have an effect on $\lambda$. Controlling for $Q$ (i.e., keeping it fixed), the expected relation is: the smaller $\alpha$, i.e. the precision of the DPM mixture, the larger $\lambda$, i.e. the relative density around the modes; controlling for $\alpha$ (i.e., keeping it fixed), the expected relation is: the smaller $Q$, i.e. the variance of the components of the DPM, the larger $\lambda$. 
The parameters of the base measures $G_0$ have a non-negligible role in determining $\lambda$, so in this section focus is devoted to the relation between the precision parameter $\alpha$, the mixture components variance $Q$ and the index $\lambda$. Indeed, in all the study simulations the values of the other parameters involved in the DPM have been kept fixed across the scenarios. \\
\paragraph{\textit{Data generating process}}
The experimental design is as follows.
For 4 different values of $\alpha= (0.1, 1, 5, 20)$ and 2 different values of $Q = (0.1, 1.5)$ a set of independent observations $u=1,\dots,n$ are drawn from the following DPM:\\
\begin{eqnarray}
    u_i |\mu_c, Q, &\stackrel{iid}{\sim} &\displaystyle\sum_{c=1}^{R} \pi_c N(\mu_c, Q), \;\;\;i = 1,\ldots, n \nonumber\\
    \mu_c &\stackrel{iid}{\sim}& G_0 \nonumber\\
    \pi_c&=& \nu_c \prod_{l<c}(1-v_l),\;\;\textrm{where} \nonumber\\
    v_c  &\stackrel{iid}{\sim} & Be(1,\alpha), \;\;c=1\ldots,R. \nonumber
\end{eqnarray}
\\
Where $\mu_c$ and $\pi_c$ are the location parameter and the mixing proportion of the component $c$, respectively; $G_0$ is the base measure; and $\nu_c$ is the parameter of the  stick-breaking. 
Following the above mentioned truncated stick-breaking construction, here $R$ is the maximum number of observable cluster. Across the eighth scenarios the following quantities are assigned: the number of observations $n=500$, the maximum number of clusters $R=50$, the base measure $G_0:$ $U(-6,6)$. Here, $U(\cdot)$ stands for uniform distribution. The use of these distributions in the present experimental context aims to highlight the effect of different values of $\alpha$ and $Q$ on $\lambda$ in a more evident and interpretable manner. \\

\subsubsection{Results}
As shown in Tables \ref{table:1} and \ref{table:2} as $\alpha$ increases, and so the number of point masses of $G$ increases as well, $\lambda$ decreases. The density of the observations $u=1,\dots,n$ is concentrated around few point masses (few modes) for lower value of $\alpha$ and is spread out the larger. Note also the change of density of the antimodes. As expected, it is proportional to the precision parameter in a positive fashion. As the observations are more spread as $\alpha$ increases, there are fewer intervals in the support with relative small density: $\lambda$ index decreases at larger values of $\alpha$ (column-wise Table \ref{table:1} and Table\ref{table:2}). A similar proportional relation is observed between the variance of the mixture components $Q$ and $\lambda$ when $\alpha$ is kept fixed (row-wise Table \ref{table:1} and Table\ref{table:2}). Smaller values of both $\alpha$ and $Q$ result in a high polarized distribution of $u=1,\dots,n$ and correspond to larger values of $\lambda$. Whereas larger values of both $\alpha$ and $Q$ result in a low polarized distribution and correspond to smaller values of $\lambda$. 
It is is an index of how spread the density is over the support of the hierarchical effects.\\
From an interpretative point of view, $\lambda$ indicates the degree of overlap between the infinitely many clusters. It might be informative of the separation between them. Since this quantification is based on a non-parametric density, $\lambda$ is not directly related to the number of the group, or to the cluster location. It indicates the degree to which the independent observations drawn from a DPM overlap; the variance of the cluster $Q$ also plays a crucial role. The index thus quantifies the combined effect of the parameters to assess the extent to which possible different opinions (i.e., the modes) might be strongly shared among the raters (i.e., the modes are sharp pick of density). To this regard, $\lambda$ is a polarization index in presence of heterogeneity. The higher the polarization levels, the larger the values of the index. The practical interpretation and the operational decisions must be guided by the field of application.

 \begin{table}
        \centering
        \begin{tabular}{cM{5.3cm}M{4.8cm}M{4.8cm}}
           \toprule
            & $Q_c = 0.1$ &  $Q_c = 1.5$\\
            \midrule\\
            $\alpha = 0.1$ & \includegraphics[scale=0.26]{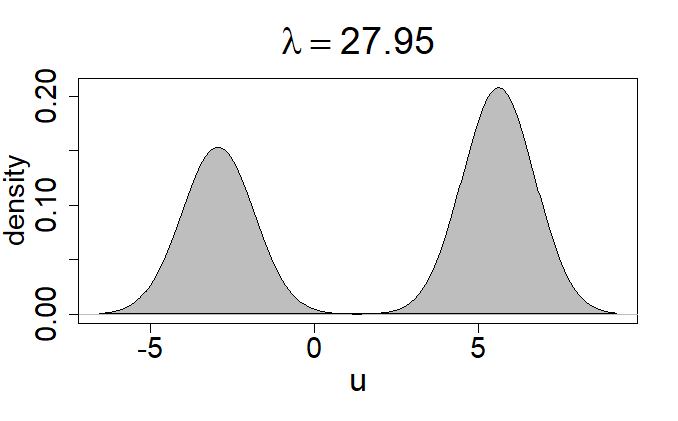} & \includegraphics[scale=0.26]{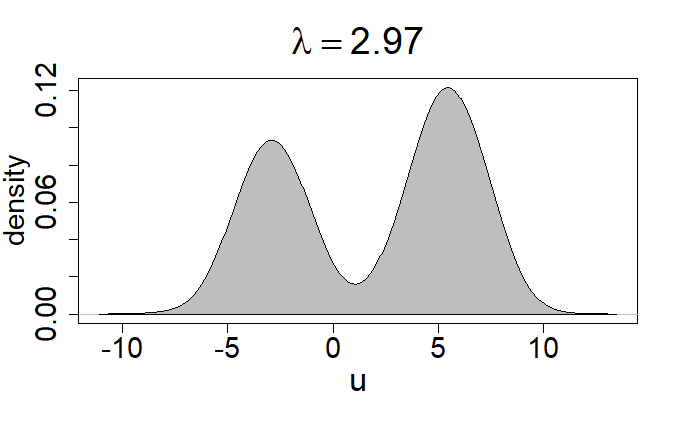}  \\
           $\alpha = 1$ & \includegraphics[scale=0.26]{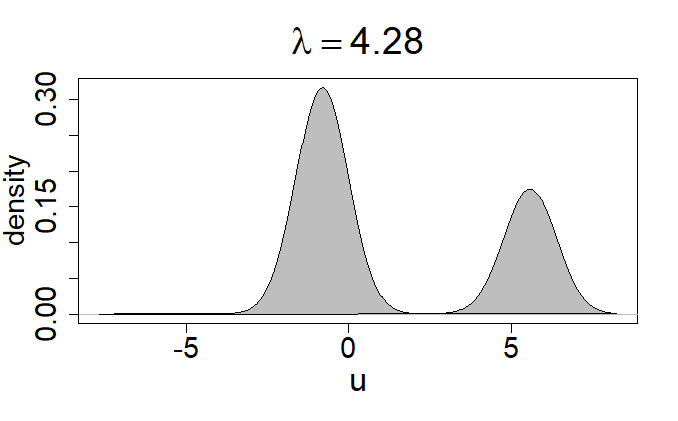} & \includegraphics[scale=0.26]{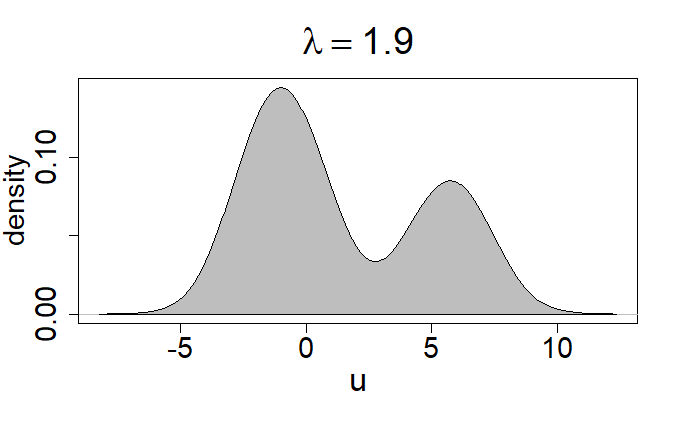} \\
            $\alpha = 5$ & \includegraphics[scale=0.26]{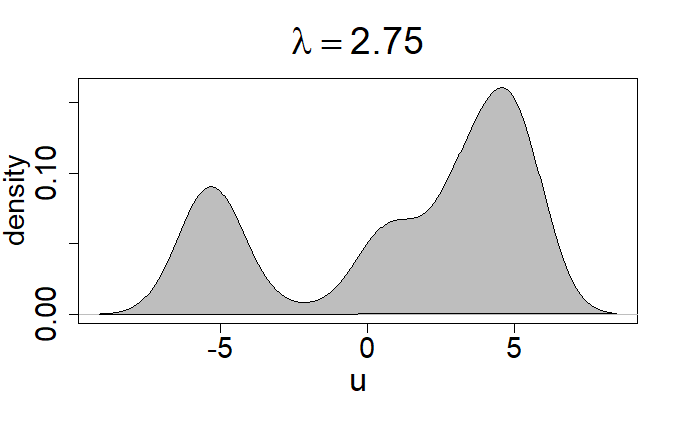} & \includegraphics[scale=0.26]{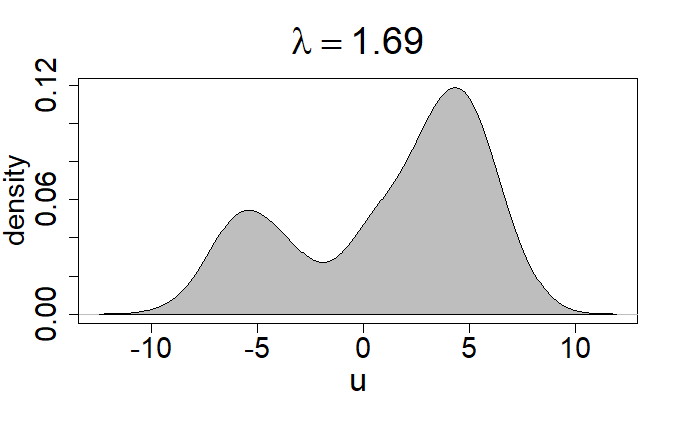} \\
           $\alpha = 20$ & \includegraphics[scale=0.26]{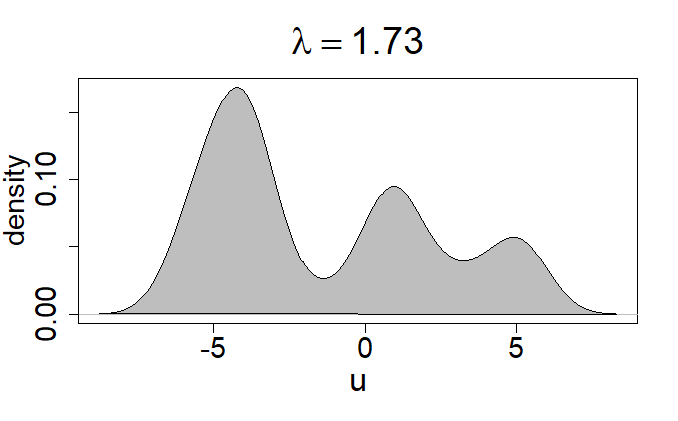} & \includegraphics[scale=0.26]{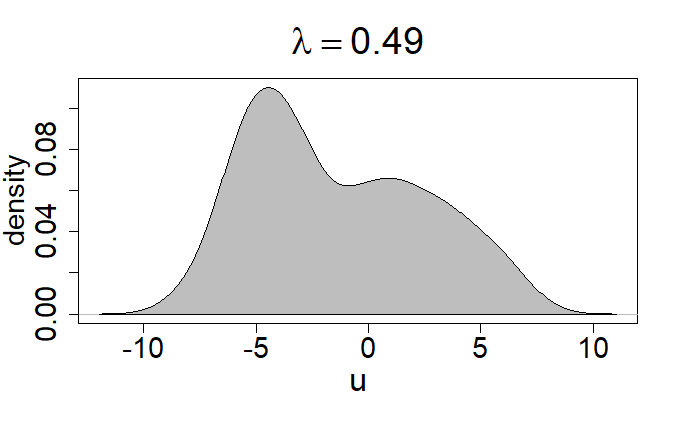} \\
            \bottomrule
        \end{tabular}
        \caption{The eight scenarios correspond to a DPM with different values of the precision parameter $\alpha$ and the components variance $Q$. Each scenario correspond to a specific combination of these two parameters. All the other quantities are fixed across the scenarios. The realizations of the DPM are here indicated as $u= 1,\dots,n$. Different combinations of $\alpha$ and $Q$ result in different values of $\lambda$. For fixed values of $Q$ (column-wise), a proportional relation is shown between $\alpha$ and $\lambda$: when the first increases, the second decreased. Similarly, for fixed values of $\alpha$ (row-wise), a proportional relation is shown between $Q$ and $\lambda$: when the first increases, the second decreased. Smaller values of both $\alpha$ and $Q$ result in a high polarized distribution of $u=1,\dots,n$ and correspond to larger values of $\lambda$. Whereas larger values of both $\alpha$ and $Q$ result in a low polarized distribution and correspond to smaller values of $\lambda$. }
        \label{table:1}
    \end{table}

\begin{center}
  \begin{table}[H]
      \centering
      \begin{tabular}{ |c|c|c|c| }
\hline
      ~  {} ~ &   ~~~~ {$\alpha$}  ~~~~ & ~~~~{$Q$} ~~~~ &   ~~~~{$\lambda$}  ~~~~\\
     \hline
        Scenario 1  &  0.1  & 0.1 &  27.95\\ \hline
       Scenario 2  &  0.1  & 1.5& 2.97 \\ \hline
       Scenario 3  &  1    & 0.1 & 4.28  \\ \hline
     Scenario 4 &   1    & 1.5& 1.9  \\ \hline
     Scenario 5 &   5    & 0.1&  2.75\\ \hline
         Scenario 6  &  5    & 1.5& 1.69 \\ \hline
         Scenario 7  &  20   & 0.1& 1.73  \\ \hline
         Scenario 8 &   20   & 1.5& 0.49  \\ 
  \hline
\end{tabular} 
      \caption{Parameters values at each scenarios. Each of them correspond to a DPM with different values of the precision parameter $\alpha$ and the components variance $Q$. Both $\alpha$ and $Q$ have an effect on the distribution polarization of the realizations of the DPM. As a result different values of $\lambda$ are observed. }
      \label{table:2}
  \end{table}
\end{center}

\vspace{2.5px}

\subsection{Simulation 2: Inter-rater agreement and $\lambda$}
\label{sub:2}
\subsubsection{Simulation setting}
The following simulation study aims to highlight the complementary role of $\lambda$ as an additional summary metric in inter-rater agreement analysis. The varying intercept parametrization is hereafter adopted as univariate case for the raters effects. To this aim the standard modelling approach (i.e., the normal distributed varying intercept and the resulting ICC) is compared with the nonparametric proposed above (i.e., the DPM prior over the varying intercept and $\lambda$). 
The experiment evaluates the performance of both standard $ICC$ and $\lambda$ in the presence of heterogeneity between raters' evaluations due to a multimodal distribution of the hierarchical effects.\\ 
\paragraph{\textit{Data generating process}}
Three experimental scenarios were planned, in which a different clustering on the raters' intercept parameter was specified in the generative model. In each scenario the rater's intercept $u_i$ was generated from a bimodal Gaussian mixture. The location parameters of the mixture components were fixed across the scenarios, $\mu_1=-3$ and $\mu_2=3$;whereas decreasing values $(1,0.5,0.1)$ were assigned to the components scale parameters $Q_1$ and $Q_2$ (see Table \ref{tab:3}). This resulted in different polarization scenarios. The mixture components were kept equiprobable ($\pi_c=0.5$), $c \in \{1,2\}$ throughout.
The number of raters $I=100$ and the number of items $J=250$ were fixed across the scenarios. One continuous covariate $x_{ij}$ with an effect $\beta=2$ was used and it was the same across the scenarios.  
\begin{center}
  \begin{table}[H]
      \centering
      \begin{tabular}{ |c|c| }
  \hline
  Scenario 1 & $u_i \stackrel{iid}{\sim} 0.5\cdot N(-3,1)+0.5\cdot N(3,1) $ \\
  \hline
  Scenario 2 & $u_i \stackrel{iid}{\sim} 0.5\cdot N(-3,0.5)+0.5\cdot N(3,0.5) $ \\
  \hline
  Scenario 3 & $u_i \stackrel{iid}{\sim} 0.5\cdot N(-3,0.1)+0.5\cdot N(3,0.1) $ \\
  \hline
\end{tabular} 
      \caption{True raters' hierarchical effects distribution across different scenarios. A Gaussian mixture is specified as distribution of the hierarchical effects $u_i=1,\dots,I$. The location parameters of two components of the mixture are kept fixed across the scenarios and decreasing values were assigned to the respective scale parameters.}
      \label{tab:3}
  \end{table}
\end{center}

\paragraph{\textit{Standard model approach}}
The following priors were specified for the standard hierarchical effect model (i.e., the varying intercepts are assumed to be i.i.d. normal distributed): 
\begin{eqnarray}
    \beta &\sim& N(0, 5), \nonumber\\
    \sigma_\epsilon &\sim& Exp(0.2), \nonumber\\
    \sigma_u &\sim& Exp(0.2), \nonumber\\
    u_i &\stackrel{iid}{\sim}& N(0,\sigma_u),\nonumber   
\end{eqnarray}
for $i =1,...,I$; $Exp(\cdot)$ stands fro the exponential distribution and $\beta$ is the non-hierarchical effect, $\sigma_u$ and $\sigma_\epsilon$ are the hierarchical effect and the noise variances parameters, respectively. A logic of complexity penalization was used in the choice of the above mentioned priors distributions \citep{Simpson_2017}. The posterior of each standard hierarchical effect model were sampled using  NUTS-Hamiltonian MCMC in Stan language \citep{Rstan}.
\paragraph{\textit{Nonparametric model approach}}
The set of priors introduce in section \ref{sec:4} were elicited for the DPM models with the following hyperparameters as suggested by \citep{Heinzl_2012}: $\mathbf{b}_0= \mathbf{0}, \mathbf{S}_0= 1000\mathbf{I}_p, a_{\beta_0}=0.005, b_{\beta_0}=0.005, \mathbf{m}_0= \mathbf{0}, \mathbf{W}_0=100\mathbf{I}_q, a_{D_0}=0.5, b_{D_0}=0.5, a_{Q_0}=0.001, b_{Q_0}=0.001, a_\alpha=2 , b_\alpha=2 ,a_\epsilon=0.005, b_\epsilon=0.005 $. As result of some preliminary analysis, a dense grid of 481 equally-spaced values from -12 to 12 (i.e., with a fixed interval of 0.05) was used to monitoring the mixture density of the nonparametric varying intercept $u_i$ at each iteration. The posterior distribution of the nonparametric hierarchical effect is obtained as the set of the mean density of each point of the grid over the iterations \citep{Gelman2013}.  \\ 
In all the computations for both the models 55,000 iteration with 5,000 burn-in were used, the  Markov chains were thinned the by a factor of 50, resulting in samples of size 1000 \citep{Heinzl_2012}. \\

\subsubsection{Results}
\begin{figure}
    \centering
    \subfigure[]{\includegraphics[scale=0.27]{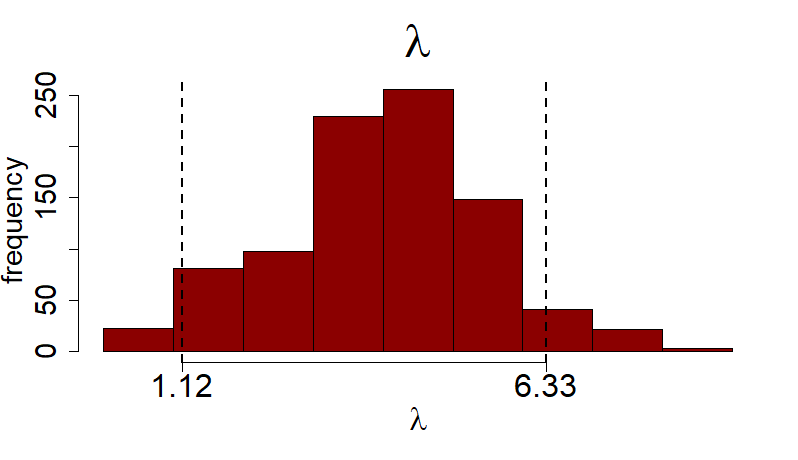}} 
    \subfigure[]{\includegraphics[scale=0.27]{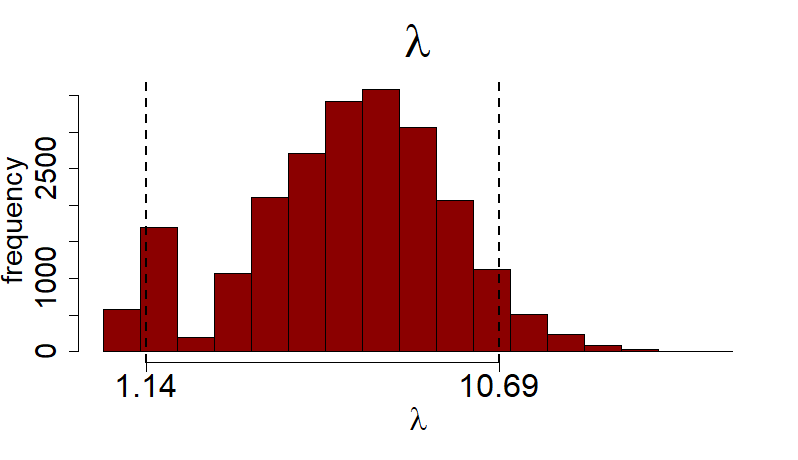}} 
    \subfigure[]{\includegraphics[scale=0.27]{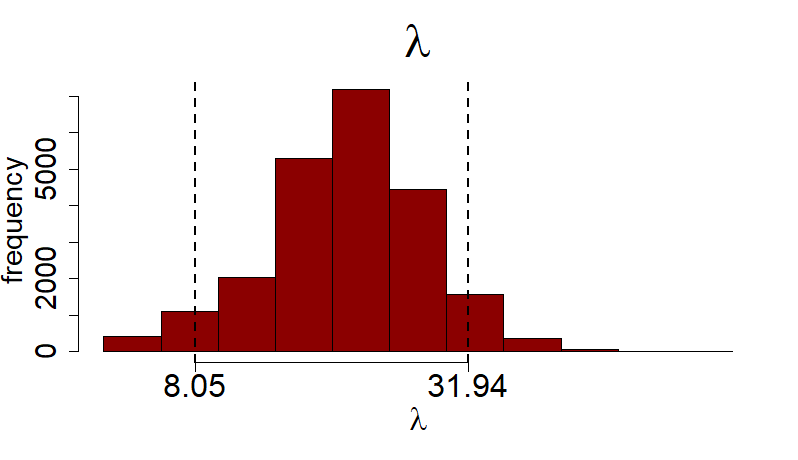}}
    \caption{Posterior distribution of $\lambda$: (a) Scenario 1: lower polarization (b) Scenario 2: medium polarization (c). Scenario 3: higher polarization. The black dotted lines stands for 95\% credible intervals.}
    \label{fig:4}
\end{figure}

\begin{figure}
    \centering
    \subfigure[]{\includegraphics[scale=0.27]{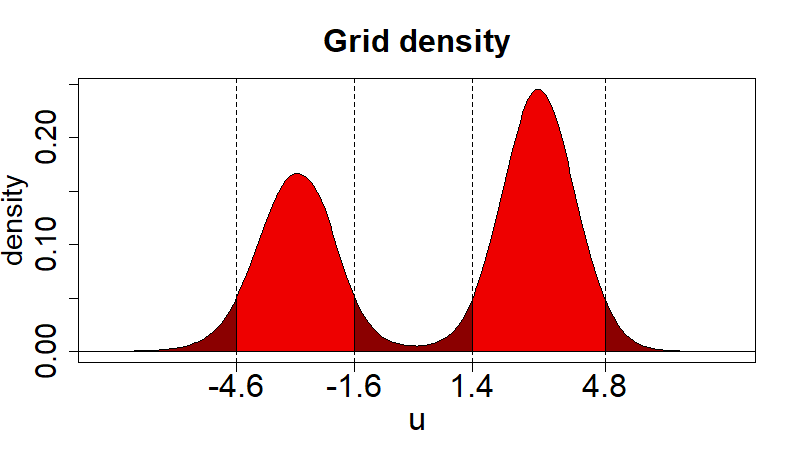}} 
    \subfigure[]{\includegraphics[scale=0.27]{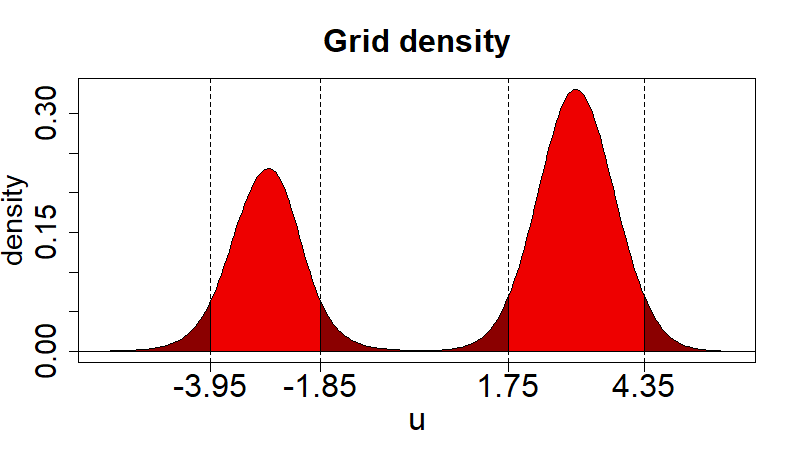}} 
    \subfigure[]{\includegraphics[scale=0.27]{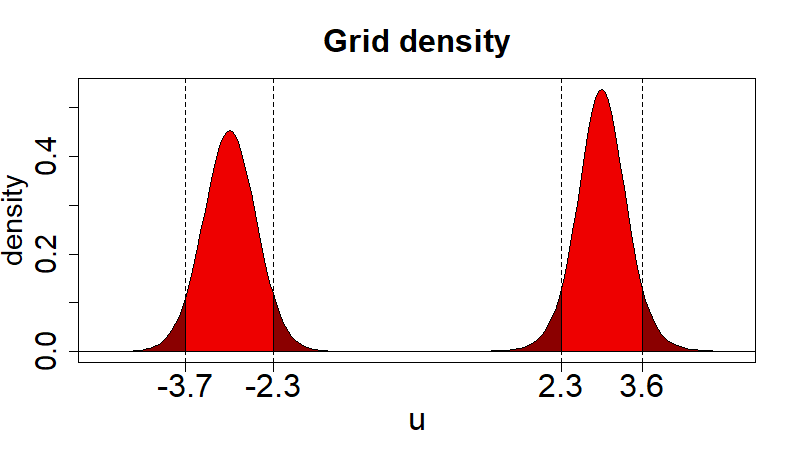}}
    \caption{95\% HPD intervals of the hierarchical effect from the DPM. The different mixture used to generate the data emerged clearly from the posterior of the grid adopted to monitoring $\mathbf{u}$. (a) Scenario 1 (b) Scenario 2 (c). Scenario 3}
    \label{fig:2}
\end{figure}

\begin{figure}
    \centering
    \subfigure[]{\includegraphics[scale=0.27]{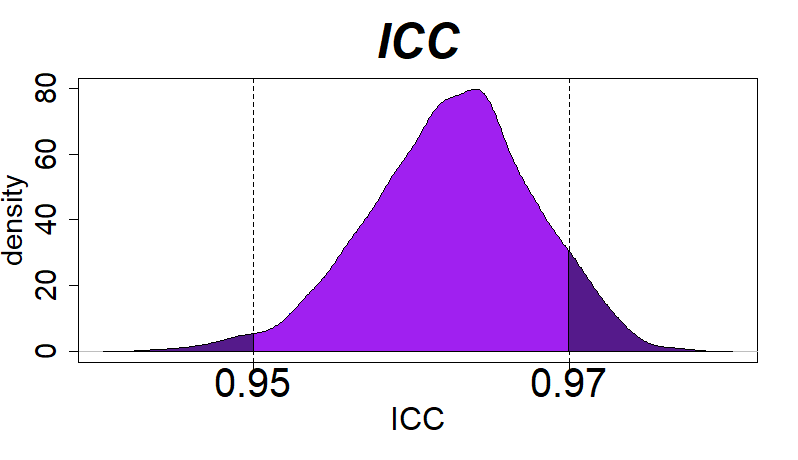}} 
    \subfigure[]{\includegraphics[scale=0.27]{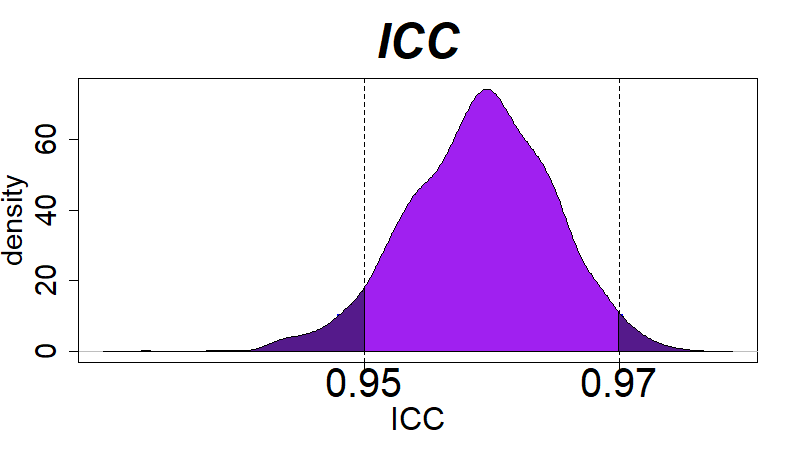}} 
    \subfigure[]{\includegraphics[scale=0.27]{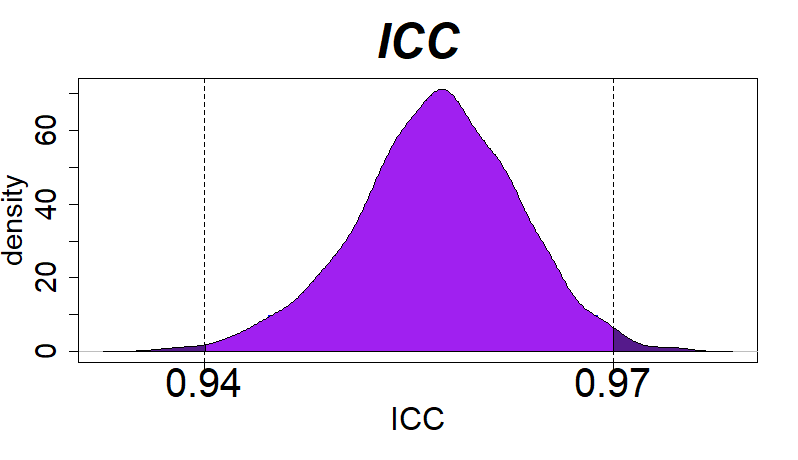}}
    \caption{95\% HPD intervals of ICC from the the standard models (i.i.d. normal distributed varying intercepts): (a) Scenario 1: lower polarization (b) Scenario 2: medium polarization (c). Scenario 3: higher polarization.}
    \label{fig:5}
\end{figure}

As shown in Table \ref{table:4} the standard model (i.e., that in which hierarchical raters intercepts are assumed to be i.i.d. normally distributed) due to the rigid distributional assumption of the hierarchical parameters is not able to capture the possible multimodal distribution and it resulted in a large value of the hierarchical effect variance $\sigma_u$ (see  Table \ref{table:4} and Figure \ref{fig:2} ). As a result, the ICC didn't capture almost any difference among the three different scenarios (see Table \ref{table:4} and Figure \ref{fig:5}). On the contrary, the DPM model, due to the flexible nonparametric specification of the intercepts prior, showed a good performance. As evident from Figure \ref{fig:2} and Table \ref{table:4} the DPM model was far more able to reproduce the data generating process. The different mixture used to generate the data emerged clearly from the posterior of the grid adopted to monitoring $\mathbf{u}$. Since the DPM model properly learn the multi-modalities of the raters intercepts density, the index $\lambda$, being  based on the ratio between the mean density of the modes and that of the antimodes present in the grid at each iteration, showed to be able to differentiate the three different polarization scenarios. It gives some interesting information regarding the shape of the non-parametric mixture distribution. The 95\% credible interval of $\lambda$ (see Table \ref{table:5} and Figure\ref{fig:4}) as estimated in the three different scenarios highlighted different degrees of amplitude and separation (i.e., different degrees of polarization) along them. The index $\lambda$ is computed as a logarithm of the ratio of the average mode density against the density of the antimodes at each iteration $t$ of the posterior sampler. So, in the first scenario, the values of the 95\% credible interval are smaller, indicating that in most of the iterations the difference between the mean density at the modes and that of the antimodes was very small. In terms of the third scenario, $\lambda$ assumed rather larger values along the iterations as evidence that the mode density is far larger than that of the antimodes. In other words, the rater clusters were separated and  clearly distinct. The parameters of the DPM are the most influential with regard to $\lambda$. Specifically, the location parameters $\mu_c$, $c=1,\dots,R$, and the scale parameters $Q_c$, $c=1,\dots,R$, showed to have a combined effect of the proposed index.
The 95\% HDP intervals of the parameters of both the DPM prior model and that with normal distributional assumption are reported in tables \ref{table:6} and \ref{table:7}, respectively. \\

\begin{table}[H] 

\centering
\begin{tabular}{|c|c|c|}
\hline
         &$\sigma_u$  & Grid density \\ \hline
         Scenario 1 & $(2.80, 3.75)$ & $(-4.60, -1.60) \cup (1.40, 4.60) $\\ \hline
         Scenario 2 & $(2.65, 3.55)$ & $(-3.95, -1.85) \cup (1.75, 4.35) $ \\ \hline
         Scenario 3 & $(2.57, 3.46)$ & $(-3.70, -2.30) \cup (2.30, 3.60)$  \\ \hline
\end{tabular}
\caption{95\% HPD intervals of the hierarchical effects variance $\sigma_u$ from the standard models (i.i.d. normal distributed varying intercepts) and of the Grid density of the hierarchical effect in DPM models.} 
\label{table:4} 
\end{table}
\begin{table}[H] 
\centering
\begin{tabular}{|c|c|c|}
\hline
         & $ICC$ & $\lambda$ \\ \hline
         Scenario 1& $(0.952, 0.973)$ & $(1.22, 6.33)$ \\ \hline
         Scenario 2& $(0.948, 0.970)$ & $(1.14, 10.69)$\\ \hline
         Scenario 3 & $(0.945, 0.969)$ & $(0.05, 31.94)$ \\ \hline
\end{tabular}
\caption{95\% HPD intervals of the ICC from the standard models (i.i.d. normal distributed varying intercepts) and of $\lambda$ from the DPM models.
}
\label{table:5} 
\end{table}

\begin{table}[H] 
\centering
\begin{tabular}{|c|c|c|c|}
\hline
         & Scenario 1 &  Scenario 2 & Scenario 3 \\ \hline
         $\beta$& $(1.87,2.12)$ & $(1.87, 2.13)$ & $(1.87,2.12)$ \\ \hline
          $b_\beta$& $(-7.15,11.03)$ & $(-7.26, 10.58)$ & $(-7.44, 10.97)$ \\ \hline
         $\sigma_\beta$& $(0.11,91.40)$ & $(0.11, 88.28)$ & $(0.10,87.69)$ \\ \hline
        $\mu_0$& $(-0.16,1.16)$ & $(0.12,1.33)$ & $(-0.49,0.70)$\\ \hline
         $\sigma_{D_0}$ & $(6.63,12.63)$ & $(6.09,11.30)$ &$(6.52,11.68)$ \\ \hline
         $\sigma_\epsilon$ & $(0.43,0.46)$ & $(0.42,046)$ & $(0.42,0.46)$ \\ \hline
         $\alpha$ & $(8.05,18.06)$ & $(8.15,18.09)$ & $(8.12,18.03)$ \\ \hline
\end{tabular}
\caption{95\% crebible intervals of $\beta$, DPM and residuals related parameters. Here $\beta$ is the non varying effect, $b_\beta$ and $\sigma_\beta$ are, respectively, the related location and scale hyperparameters; $\mu_0$ and $\sigma_{D_0}$ are the location and scale parameters of the base measure $G_0$, respectively. The precision parameter $\alpha$ and the residuals standard deviation $\sigma_\epsilon$ are also reported.  
}
\label{table:6} 
\end{table}

\begin{figure}
    \centering
    \subfigure[]{\includegraphics[scale=0.27]{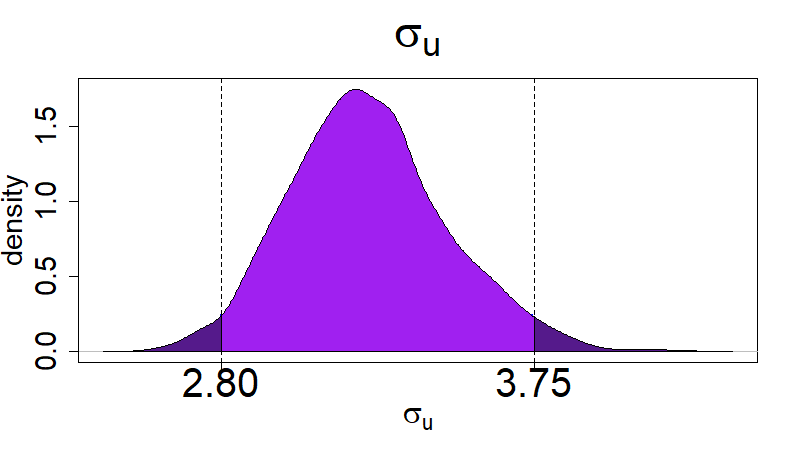}} 
    \subfigure[]{\includegraphics[scale=0.27]{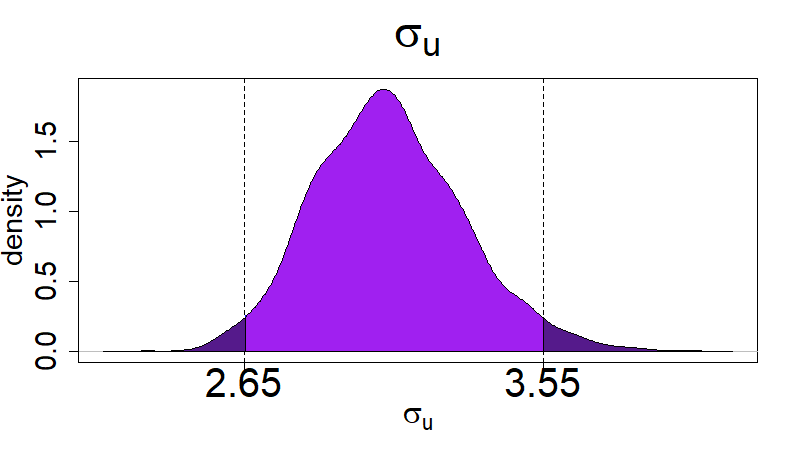}} 
    \subfigure[]{\includegraphics[scale=0.27]{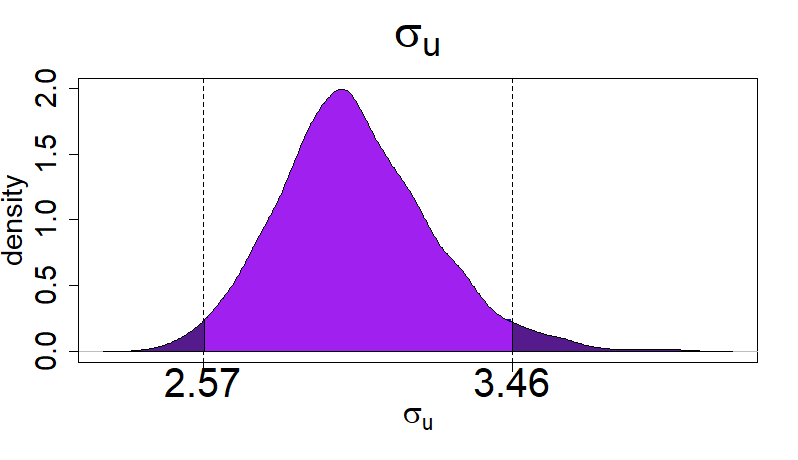}}
    \caption{95\% HPD intervals of $\sigma_u$ from the standard models (i.i.d. normal distributed varying intercepts): (a) Scenario 1: lower polarization (b) Scenario 2: medium polarization (c). Scenario 3: higher polarization. Scenario 3: higher polarization. These models, due to their rigid distributional assumption, poorly differentiate the three different polarization scenarios.}
    \label{fig:3}
\end{figure}
\vspace{2.5px}

\vspace{2.5px}

\vspace{2.5px}

\vspace{2.5px}

\begin{table}[H]
\centering
\begin{tabular}{|c|c|c|c|}
\hline
         & Scenario 1 & Scenario 2 &Scenario 3 \\ \hline
         $\beta$ & $(1.99, 2.00)$ & $(1.98, 2.01)$ & $(1.99, 2.01)$\\ \hline
         $\sigma_\epsilon$ & $(0.60, 0.64)$ & $(0.62, 0.63)$ & $(0.62, 0.63)$ \\ \hline
\end{tabular}
\caption{95\% HPD intervals of the other parameters of the standard models (i.i.d. normal distributed varying intercepts)}
\label{table:7} 
\end{table}

\section{Large scale performance assessment}
The evaluation heterogeneity of teachers is a long-standing issue in psychometrics \citep{Uto_2022,Shirazi_2019, Bonefeld_2018,Casabianca_2015,DeCarlo_2008}. Highly biased scored might have a detrimental effect on students proficiency and education \citep{Chin_2020,Paredes_2014,Cooper_2003}. The proposed nonparametric model and the index $\lambda$ might be valuable tools to address this issue. They might help to shed light on very biased assessment contexts and to provide fairer scores. The estimated hierarchical effect of each teacher (which may be interpreted as the teacher's bias) might be used to adjust the observed score. The index $\lambda$ might quantify teachers polarization in their grading.\\  
\paragraph{\textit{The \textit{Matura} data set}}
As an illustrative real data application, a large scale performance assessment data set was analysed \citep{Zupanc2018}. The DPM-model was applied to a large-scale essay assessment data obtained during the nation-wide external examination conducted by the National Examination Centre in upper secondary schools in Slovenia also known as \textit{Matura} and analyzed in \cite{Zupanc2018}. These data were related to the spring term argumentative essays for years between 2010 and 2014. Particular attention is devoted to the distinction between two main aspects of essay writing: the language correctness (i.e., the presence of grammatical or syntactic errors) and the the good argumentation of the content (i.e., a good and clear presentation of all the arguments).
Regarding the data structure, students are nested within the teachers. So that each student's essay is evaluated by one trained teacher, who is asked to grade it concerning two different rubrics. An essay can receive a score between 0 and 20 for the language-related rubric and between 0 and 30 for the content-related one. 
Prior analysis of these data \citep{Zupanc2018} revealed that heterogeneity among teachers was broadly down to two types: strict and lenient. The two different trends might be captured by the model and their polarization quantified by the $\lambda$ index.\\
For this reason N=2616 students' essays, each scored by one of I=18 different teachers, were considered for the analysis \footnote{For illustrative purposes, only the variables related to the first teachers were considered.}. The objective of this application is to analyze teachers' individual differences in scoring the essay content, controlling for its language correctness. How lenient or strict they are in scoring the quality of an essay content, without the effect of the language correctness. 
The content score is commonly ways more susceptible to idiosyncrasies or biases of the teacher than the language-related score, which is generally more objective \citep{Tasha_2023,Zhu_2021,Shirazi_2019}. 
Accordingly, the content-related score was specified as outcome variable and the language-related score as covariate with a non varying  effect. 
A DPM hierarchical prior was specified over the teachers' intercepts.
All the scores were re-scaled for this analysis to get a easier parameters value interpretation \footnote{The following transformation was applied to standardize the score: $f(x)=\frac{x-\overline{x}}{\hat{\sigma_x}}$, where $\overline{x}=\frac{1}{N}\sum_{n=1}^{N}x_n$ was the sample mean and $\hat{\sigma_x}=\sqrt{\frac{1}{N}\sum_{n=1}^{N-1}(x_n-\overline{x})^2}$ the sample standard deviation.} \citep{Gelman2013}. 

\subsection{Results}
\begin{figure}
    \centering
    \subfigure[]{\includegraphics[scale=0.27]{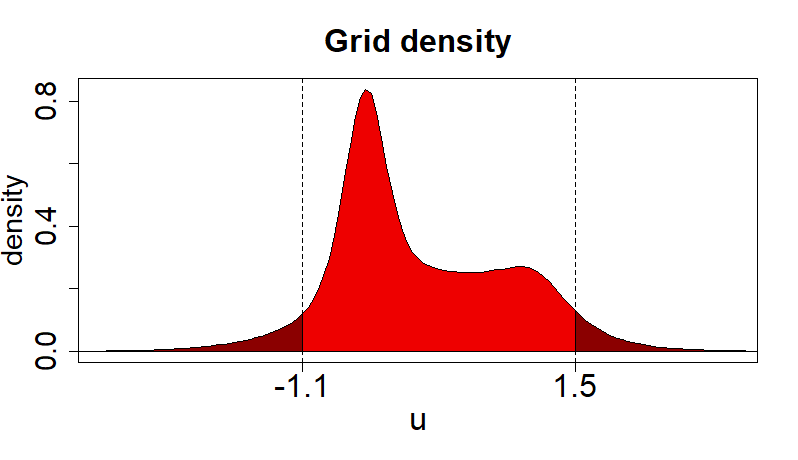}} 
    \subfigure[]{\includegraphics[scale=0.27]{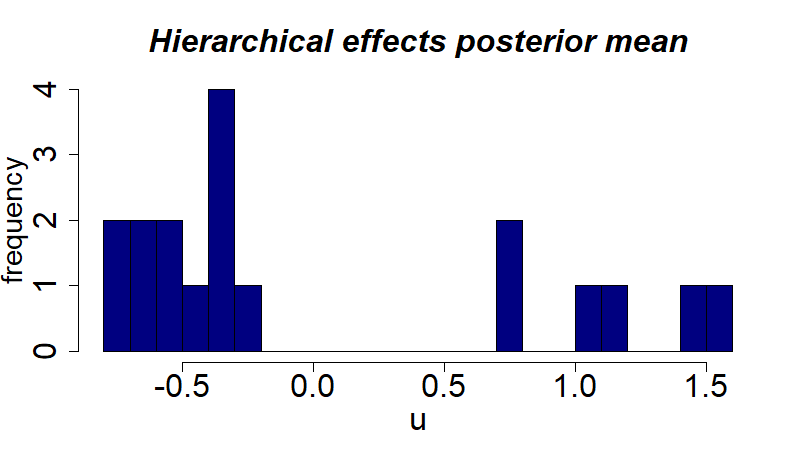}} 
    \subfigure[]{\includegraphics[scale=0.27]{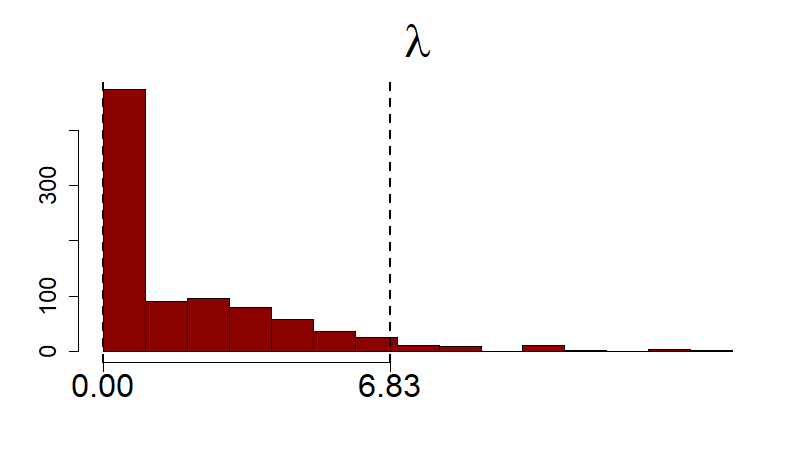}}
    \caption{(a) 95\% HPD of the monitoring grid for the teachers' hierarchical effects $u=1,\dots,I$ (b) Posterior mean of the hierarchical effect of each teacher. Two different clusters emerged from the analysis, as expected: the more lenient (to the right-hand side) and the stricter (to the left-hand side). (c) $\lambda$' 95\% credible intervals. It indicate a moderate polarized posterior distribution of the posterior hierarchical effects. }
    \label{fig:6}
\end{figure}

The language-related score showed a posterior mean effect of 0.27 on the content-related score, with a (0.16, 0.38) 95\% credible interval. The language correctness of the essay writing had moderate role in predicting the evaluation of the its content. 
As shown by Figure \ref{fig:6}(a) the DPM-model learned the presence of two main trends from the data. The bimodal non-parametric distribution over the grid suggested that the teachers were rather heterogeneous in the essay scoring process. More precisely, they seemed to be slightly polarized around two main tendencies. Some teacher showed a slightly more lenient or stricter than the others (i.e., who had a larger or smaller hierarchical effect posterior mean, respectively), see Figure \ref{fig:6}.  The $\lambda$ index showed a posterior mean of 1.87 which suggested a low polarization. The $\lambda$ 95\% HPD interval was (0.0, 6.83) which indicated a non negligible occurrence of quite high values of $\lambda$. All the other parameters credible intervals are reported in Table \ref{table:8}\\
Assuming this latent group polarization as a low latent agreement among raters, the $\lambda$ index might be used in a diagnostic manner. Considering the present application, some solutions might be suggested for a fairer assessment process. Firstly, assuming a very negligible noise term, the teacher's estimated bias might be removed from the actual score. Another practical solution might be the implementation \textit{ad hoc} training aimed to a much more shared point of view in essay scoring.   

\begin{table}[H] 
\centering
\begin{tabular}{|c|c|}
\hline
         $\beta$& $(0.16,0.38)$ \\ \hline
          $b_\beta$& $(-14.80,20.32)$  \\ \hline
         $\sigma_\beta$& $(0.11,98.33)$  \\ \hline
         $\mu_0$& $(-0.33,0.34)$ \\ \hline
         $\sigma_{D_0}$ & $(0.19, 1.03)$ \\ \hline
         $\sigma_\epsilon$ & $(0.13,0.15)$  \\ \hline
         $\alpha$ & $(3.00,4.17)$  \\ \hline
\end{tabular}
\caption{95\% credible intervals of $\beta$, DPM and residuals related parameters. Here $\beta$ is the non varying effect, $b_\beta$ and $\sigma_\beta$ are, respectively, the related location and scale hyperparameters; $\mu_0$ and $\sigma_{D_0}$ are the location and scale parameters of the base measure $G_0$, respectively. The precision parameter $\alpha$ and the residuals standard deviation $\sigma_\epsilon$ are also reported.  
}
\label{table:8} 
\end{table}

\section{Conclusions}
 \label{sec:7}   
Most of the statistical models commonly used to analyze data from such observational contexts haven't shown to be very flexible to certain types of heterogeneity among raters. The common HLMs with a normal (or unimodal) distributional assumption for the hierarchical effects cannot capture any possible latent clusters, i.e. any multimodality. Indeed, the residual covariance modelled through the hierarchical effects might be informative about different latent similarities among raters. In this regard, incorporating a DPM in the prior of the hierarchical effects distribution is a flexible  choice to address this issue. \\
Consequently, the estimation of the agreement among the raters should take into account the possible multimodal distribution of the hierarchical effects. Interest might not be exclusively on the proportion of variance attributable to the hierarchical effects over the total variance (i.e., the main interpretation of the ICC); instead, it might be more appealing to explore the entire multimodal density. Since the DPM naturally accommodates clusters among hierarchical effects (i.e., among raters), it is natural to consider the extent to which the mixture components are separated. Since $\lambda$ is based on the density approximated through the grid approach $f(\mathbf{u})$, it reflects both the clustering induced by the Dirichlet process and the variance of the mixture components. 
Due to the particular information carried by $\lambda$ it might be more informative about the latent agreement among raters than the solely ICC. The latter is very useful when the normal distributional assumption of the hierarchical effects holds. However, in the presence of multimodality the estimate of the variance of the hierarchical effect $\sigma_u$ is not accurate (it might be over-estimated) and the related ICC might be non-informative. \\
In contexts in which strong beliefs about the exact number of cluster are present or it is supported by some sort of evidence, an hierarchical model with a prior finite mixture distribution over the hierarchical effects is expected to have comparably good performance as well.  The parametric variance of a mixture might be take into account in the ICC formula in these cases. For the above mentioned reasons, added flexibility and the shrinkage property the DPM was here preferred. \\
Many other studies are needed to fully understand the performance of $\lambda$ across different combinations of the Dirichlet process parameters. Future works might highlight the role of $\lambda$ when the rating is either expressed on a dichotomous or on a polytomous scale. Further studies might highlight the computation of $\lambda$ when multivariate hierarchical effects are specified.
Comparisons between this index and the others widely used in these cases \cite{Tang2022,Forchheimer_2015,Zang2003} might be a focus of future studies. Further application of $\lambda$ in a non-parametric context might be studied \citep{Canale_2017}.

\bibliography{references}

\begin{thebibliography}{}

\bibitem[Agresti, 2015]{Agresti2015}
Agresti, A. (2015).
\newblock {\em Foundations of Linear and Generalized Linear Models}.

\bibitem[Antoniak, 1974]{Antoniak_1974}
Antoniak, C.~E. (1974).
\newblock {Mixtures of Dirichlet Processes with Applications to Bayesian Nonparametric Problems}.
\newblock {\em The Annals of Statistics}, 2(6):1152 -- 1174.

\bibitem[Barneron et~al., 2019]{Barneron_2019}
Barneron, M., Allalouf, A., and Yaniv, I. (2019).
\newblock Rate it again: Using the wisdom of many to improve performance evaluations.
\newblock {\em Journal of Behavioral Decision Making}, 32(4):485--492.

\bibitem[Bartoš et~al., 2020]{Martinkova}
Bartoš, F., Martinkova, P., and Brabec, M. (2020).
\newblock {\em Testing Heterogeneity in Inter-Rater Reliability}, pages 347--364.

\bibitem[Blackwell, 1973]{Blackwell_1973}
Blackwell, D. (1973).
\newblock {Discreteness of Ferguson Selections}.
\newblock {\em The Annals of Statistics}, 1(2):356 -- 358.

\bibitem[Blackwell and MacQueen, 1973]{Blackwell_MacQueen}
Blackwell, D. and MacQueen, J.~B. (1973).
\newblock {Ferguson Distributions Via Polya Urn Schemes}.
\newblock {\em The Annals of Statistics}, 1(2):353 -- 355.

\bibitem[Bonefeld and Dickhäuser, 2018]{Bonefeld_2018}
Bonefeld, M. and Dickhäuser, O. (2018).
\newblock (biased) grading of students’ performance: Students’ names, performance level, and implicit attitudes.
\newblock {\em Frontiers in Psychology}, 9.

\bibitem[Bouchard-Côté et~al., 2017]{Bouchard-Côté_2017}
Bouchard-Côté, A., Doucet, A., and Roth, A. (2017).
\newblock Particle gibbs split-merge sampling for bayesian inference in mixture models.
\newblock {\em Journal of Machine Learning Research}, 18:1--39.

\bibitem[Briesch et~al., 2014]{Briesch_2014}
Briesch, A., Hemphill, E., Volpe, R., and Daniels, B. (2014).
\newblock An evaluation of observational methods for measuring response to classwide intervention.
\newblock {\em School psychology quarterly : the official journal of the Division of School Psychology, American Psychological Association}, 30.

\bibitem[Bygren, 2020]{Bygren_2020}
Bygren, M. (2020).
\newblock Biased grades? changes in grading after a blinding of examinations reform.
\newblock {\em Assessment \& Evaluation in Higher Education}, 45(2):292--303.

\bibitem[Canale and Dunson, 2011]{Canale_2011}
Canale, A. and Dunson, D.~B. (2011).
\newblock Bayesian kernel mixtures for counts.
\newblock {\em Journal of the American Statistical Association}, 106(496):1528--1539.
\newblock PMID: 22523437.

\bibitem[Canale and Prünster, 2017]{Canale_2017}
Canale, A. and Prünster, I. (2017).
\newblock Robustifying bayesian nonparametric mixtures for count data.
\newblock {\em Biometrics}, 73(1):174--184.

\bibitem[Cao et~al., 2010]{Cao_2010}
Cao, J., Stokes, S.~L., and Zhang, S. (2010).
\newblock A bayesian approach to ranking and rater evaluation: An application to grant reviews.
\newblock {\em Journal of Educational and Behavioral Statistics}, 35(2):194--214.

\bibitem[Casabianca et~al., 2015]{Casabianca_2015}
Casabianca, J.~M., Lockwood, J.~R., and Mccaffrey, D.~F. (2015).
\newblock Trends in classroom observation scores.
\newblock {\em Educational and Psychological Measurement}, 75:311--337.

\bibitem[Childs and Wooten, 2023]{Tasha_2023}
Childs, T.~M. and Wooten, N.~R. (2023).
\newblock Teacher bias matters: an integrative review of correlates, mechanisms, and consequences.
\newblock {\em Race Ethnicity and Education}, 26(3):368--397.

\bibitem[Chin et~al., 2020]{Chin_2020}
Chin, M.~J., Quinn, D.~M., Dhaliwal, T.~K., and Lovison, V.~S. (2020).
\newblock Bias in the air: A nationwide exploration of teachers’ implicit racial attitudes, aggregate bias, and student outcomes.
\newblock {\em Educational Researcher}, 49(8):566--578.

\bibitem[Cicchetti, 1976]{cicchetti_1976}
Cicchetti, D.~V. (1976).
\newblock Assessing inter-rater reliability for rating scales: Resolving some basic issues.
\newblock {\em British Journal of Psychiatry}, 129(5):452–456.

\bibitem[Cooper, 2003]{Cooper_2003}
Cooper, C.~W. (2003).
\newblock The detrimental impact of teacher bias: Lessons learned from the standpoint of african american mothers.
\newblock {\em Teacher Education Quarterly}, 30(2):101--116.

\bibitem[Crimmins et~al., 2016]{Crimmins_2016}
Crimmins, G., Nash, G., Oprescu, F., Alla, K., Brock, G., Hickson-Jamieson, B., and Noakes, C. (2016).
\newblock Can a systematic assessment moderation process assure the quality and integrity of assessment practice while supporting the professional development of casual academics?
\newblock {\em Assessment \& Evaluation in Higher Education}, 41(3):427--441.

\bibitem[Dahlin et~al., 2016]{Dahlin_2016}
Dahlin, J., Kohn, R., and Sch{\"o}n, T.~B. (2016).
\newblock Bayesian inference for mixed effects models with heterogeneity.

\bibitem[De~la Cruz-Mesia and Marshall, 2006]{DelaCruz_2006}
De~la Cruz-Mesia, R. and Marshall, G. (2006).
\newblock Non-linear random effects models with continuous time autoregressive errors: a bayesian approach.
\newblock {\em Statistics in Medicine}, 25(9):1471--1484.

\bibitem[DeCarlo, 2008]{DeCarlo_2008}
DeCarlo, L.~T. (2008).
\newblock Studies of a latent-class signal-detection model for constructed-response scoring.
\newblock {\em ETS Research Report Series}, 2008(2):i--55.

\bibitem[Dee, 2005]{Dee_2005}
Dee, T.~S. (2005).
\newblock A teacher like me: Does race, ethnicity, or gender matter?
\newblock {\em American Economic Review}, 95(2):158--165.

\bibitem[DiMaggio et~al., 1996]{DiMaggio_1996}
DiMaggio, P., Evans, J., and Bryson, B. (1996).
\newblock Have american's social attitudes become more polarized?
\newblock {\em American Journal of Sociology}, 102(3):690--755.

\bibitem[Dorazio, 2009]{Dorazio_2009}
Dorazio, R.~M. (2009).
\newblock On selecting a prior for the precision parameter of dirichlet process mixture models.
\newblock {\em Journal of Statistical Planning and Inference}, 139(9):3384--3390.

\bibitem[Dressler et~al., 2015]{CCT2}
Dressler, W.~W., Balieiro, M.~C., and dos Santos, J.~E. (2015).
\newblock Finding culture change in the second factor: Stability and change in cultural consensus and residual agreement.
\newblock {\em Field Methods}, 27(1):22--38.

\bibitem[Esteban and Ray, 1994]{Esteban_1994}
Esteban, J.-M. and Ray, D. (1994).
\newblock On the measurement of polarization.
\newblock {\em Econometrica}, 62(4):819--851.

\bibitem[Ferguson, 1973]{Ferguson_1973}
Ferguson, T.~S. (1973).
\newblock {A Bayesian Analysis of Some Nonparametric Problems}.
\newblock {\em The Annals of Statistics}, 1(2):209 -- 230.

\bibitem[Forchheimer et~al., 2015]{Forchheimer_2015}
Forchheimer, D., Forchheimer, R., and Haviland, D. (2015).
\newblock Improving image contrast and material discrimination with nonlinear response in bimodal atomic force microscopy.
\newblock {\em Nature communications}, 6:6270.

\bibitem[Gelman et~al., 2013]{Gelman2013}
Gelman, A., Carlin, J., Stern, H., Dunson, D., and Vehtari, A.and~Rubin, D. (2013).
\newblock {\em Bayesian Data Analysis}.
\newblock Chapman and Hall/CRC.

\bibitem[Gill and Casella, 2009]{Gill_2009}
Gill, J. and Casella, G. (2009).
\newblock Nonparametric priors for ordinal bayesian social science models: Specification and estimation.
\newblock {\em Journal of the American Statistical Association}, 104(486):453--454.

\bibitem[Gisev et~al., 2013]{Gisev}
Gisev, N., Bell, J.~S., and Chen, T.~F. (2013).
\newblock Interrater agreement and interrater reliability: Key concepts, approaches, and applications.
\newblock {\em Research in Social and Administrative Pharmacy}, 9(3):330--338.

\bibitem[Gwet, 2008]{Gwet_Li}
Gwet, K.~L. (2008).
\newblock Computing inter-rater reliability and its variance in the presence of high agreement.
\newblock {\em British Journal of Mathematical and Statistical Psychology}, 61(1):29--48.

\bibitem[Heinzl et~al., 2012]{Heinzl_2012}
Heinzl, F., Kneib, T., and Fahrmeir, L. (2012).
\newblock Additive mixed models with dirichlet process mixture and p-spline priors.
\newblock {\em AStA Advances in Statistical Analysis}, 96.

\bibitem[Heinzl and Tutz, 2013]{Heinzl_2013}
Heinzl, F. and Tutz, G. (2013).
\newblock Clustering in linear mixed models with approximate dirichlet process mixtures using em algorithm.
\newblock {\em Statistical Modelling}, 13(1):41--67.

\bibitem[Hsiao et~al., 2011]{Hsiao_2011}
Hsiao, C.~K., Chen, P.-C., and Kao, W.-H. (2011).
\newblock Bayesian random effects for interrater and test–retest reliability with nested clinical observations.
\newblock {\em Journal of Clinical Epidemiology}, 64(7):808--814.

\bibitem[Ishwaran and James, 2001]{Ishwaran2001}
Ishwaran, H. and James, L. (2001).
\newblock Gibbs sampling methods for stick-breaking priors.
\newblock {\em Journal of the American Statistical Association}, 96:161--173.

\bibitem[James and Sugar, 2003]{James_2003}
James, G.~M. and Sugar, C.~A. (2003).
\newblock Clustering for sparsely sampled functional data.
\newblock {\em Journal of the American Statistical Association}, 98(462):397--408.

\bibitem[Jang et~al., 2018]{Jang}
Jang, J.~H., Manatunga, A.~K., Taylor, A.~T., and Long, Q. (2018).
\newblock Overall indices for assessing agreement among multiple raters.
\newblock {\em Statistics in Medicine}, 37(28):4200--4215.

\bibitem[Kahrari et~al., 2019]{Kahrari_2019}
Kahrari, F., Ferreira, C.~S., and Arellano-Valle, R.~B. (2019).
\newblock {Skew-Normal-Cauchy Linear Mixed Models}.
\newblock {\em Sankhya B: The Indian Journal of Statistics}, 81(2):185--202.

\bibitem[Kim et~al., 2006]{Kim_2006}
Kim, S., Tadesse, M.~G., and Vannucci, M. (2006).
\newblock {Variable selection in clustering via Dirichlet process mixture models}.
\newblock {\em Biometrika}, 93(4):877--893.

\bibitem[Kom{\'a}rek and Kom{\'a}rkov{\'a}, 2013]{Komarek_2012}
Kom{\'a}rek, A. and Kom{\'a}rkov{\'a}, L. (2013).
\newblock {Clustering for multivariate continuous and discrete longitudinal data}.
\newblock {\em The Annals of Applied Statistics}, 7(1):177 -- 200.

\bibitem[Komárek et~al., 2010]{Komarek_2010}
Komárek, A., Hansen, B.~E., Kuiper, E. M.~M., van Buuren, H.~R., and Lesaffre, E. (2010).
\newblock Discriminant analysis using a multivariate linear mixed model with a normal mixture in the random effects distribution.
\newblock {\em Statistics in Medicine}, 29(30):3267--3283.

\bibitem[Koudenburg and Kashima, 2022]{Koudenburg_2022}
Koudenburg, N. and Kashima, Y. (2022).
\newblock A polarized discourse: Effects of opinion differentiation and structural differentiation on communication.
\newblock {\em Personality and Social Psychology Bulletin}, 48(7):1068--1086.
\newblock PMID: 34292094.

\bibitem[Koudenburg et~al., 2021]{Koudenburg_2021}
Koudenburg, N., Kiers, H. A.~L., and Kashima, Y. (2021).
\newblock A new opinion polarization index developed by integrating expert judgments.
\newblock {\em Frontiers in Psychology}, 12.

\bibitem[Kyung et~al., 2011]{Kyung_2011}
Kyung, M., Gill, J., and Casella, G. (2011).
\newblock New findings from terrorism data: Dirichlet process random-effects models for latent groups.
\newblock {\em Journal of the Royal Statistical Society: Series C (Applied Statistics)}, 60(5):701--721.

\bibitem[Liljequist et~al., 2019]{Liljequist}
Liljequist, D., Elfving, B., and Skavberg~Roaldsen, K. (2019).
\newblock Intraclass correlation – a discussion and demonstration of basic features.
\newblock {\em PLOS ONE}, 14(7):1--35.

\bibitem[Lin and Lee, 2008]{Lin_2008}
Lin, T.~I. and Lee, J.~C. (2008).
\newblock Estimation and prediction in linear mixed models with skew-normal random effects for longitudinal data.
\newblock {\em Statistics in Medicine}, 27(9):1490--1507.

\bibitem[Makransky et~al., 2019]{Makransky_2019}
Makransky, G., Terkildsen, T., and Mayer, R. (2019).
\newblock Role of subjective and objective measures of cognitive processing during learning in explaining the spatial contiguity effect.
\newblock {\em Learning and Instruction}.

\bibitem[Martinková et~al., 2023]{Patrícia_2023}
Martinková, P., Bartoš, F., and Brabec, M. (2023).
\newblock Assessing inter-rater reliability with heterogeneous variance components models: Flexible approach accounting for contextual variables.
\newblock {\em Journal of Educational and Behavioral Statistics}, 48(3):349--383.

\bibitem[McCulloch and Neuhaus, 2021]{McCulloch_2021}
McCulloch, C.~E. and Neuhaus, J.~M. (2021).
\newblock Improving predictions when interest focuses on extreme random effects.
\newblock {\em Journal of the American Statistical Association}, 0(0):1--10.

\bibitem[McHugh, 2012]{McHugh12}
McHugh, M. (2012).
\newblock Interrater reliability: The kappa statistic.
\newblock {\em Biochemia medica : časopis Hrvatskoga društva medicinskih biokemičara / HDMB}, 22:276--82.

\bibitem[M{\"u}ller et~al., 2015]{Muller_2015}
M{\"u}ller, P., Quintana, F.~A., Jara, A., and Hanson, T. (2015).
\newblock {\em Bayesian nonparametric data analysis}, volume~1.
\newblock Springer.

\bibitem[Navarro et~al., 2006]{Navarro_2006}
Navarro, D.~J., Griffiths, T.~L., Steyvers, M., and Lee, M.~D. (2006).
\newblock Modeling individual differences using dirichlet processes.
\newblock {\em Journal of Mathematical Psychology}, 50(2):101--122.
\newblock Special Issue on Model Selection: Theoretical Developments and Applications.

\bibitem[Nelson and Edwards, 2015]{Nelson_2015}
Nelson, K. and Edwards, D. (2015).
\newblock Measures of agreement between many raters for ordinal classifications.
\newblock {\em Statistics in medicine}, 34.

\bibitem[Nelson and Edwards, 2008]{Nelson2008}
Nelson, K.~P. and Edwards, D. (2008).
\newblock On population‐based measures of agreement for binary classifications.
\newblock {\em Canadian Journal of Statistics}, 36.

\bibitem[Oravecz et~al., 2014]{CCT}
Oravecz, Z., Vandekerckhove, J., and Batchelder, W.~H. (2014).
\newblock Bayesian cultural consensus theory.
\newblock {\em Field Methods}, 26(3):207--222.

\bibitem[Paredes, 2014]{Paredes_2014}
Paredes, V. (2014).
\newblock A teacher like me or a student like me? role model versus teacher bias effect.
\newblock {\em Economics of Education Review}, 39:38--49.

\bibitem[Rigon and Durante, 2021]{Rigon2021}
Rigon, T. and Durante, D. (2021).
\newblock Tractable bayesian density regression via logit stick-breaking priors.
\newblock {\em Journal of Statistical Planning and Inference}, 211:131--142.

\bibitem[Rodriguez and Dunson, 2011]{Rodriguez_2011}
Rodriguez, A. and Dunson, D. (2011).
\newblock Nonparametric bayesian models through probit stick-breaking processes.
\newblock {\em Bayesian Analysis}, 6:145--178.

\bibitem[Schielzeth et~al., 2020]{Schielzeth_2020}
Schielzeth, H., Dingemanse, N.~J., Nakagawa, S., Westneat, D.~F., Allegue, H., Teplitsky, C., Réale, D., Dochtermann, N.~A., Garamszegi, L.~Z., and Araya-Ajoy, Y.~G. (2020).
\newblock Robustness of linear mixed-effects models to violations of distributional assumptions.
\newblock {\em Methods in Ecology and Evolution}, 11(9):1141--1152.

\bibitem[Sethuraman, 1994]{Sethuraman}
Sethuraman, J. (1994).
\newblock A constructive definition of dirichlet priors.
\newblock {\em Statistica Sinica}, 4(2):639--650.

\bibitem[Shirazi, 2019]{Shirazi_2019}
Shirazi, M.~A. (2019).
\newblock For a greater good: Bias analysis in writing assessment.
\newblock {\em SAGE Open}, 9(1):2158244018822377.

\bibitem[Simpson et~al., 2017]{Simpson_2017}
Simpson, D., Rue, H., Riebler, A., Martins, T.~G., and S{\o}rbye, S.~H. (2017).
\newblock {Penalising Model Component Complexity: A Principled, Practical Approach to Constructing Priors}.
\newblock {\em Statistical Science}, 32(1):1 -- 28.

\bibitem[{Stan Development Team}, 2022]{Rstan}
{Stan Development Team} (2022).
\newblock {RStan}: the {R} interface to {Stan}.
\newblock R package version 2.21.7.

\bibitem[Stefanucci and Canale, 2021]{Stefanucci2021}
Stefanucci, M. and Canale, A. (2021).
\newblock Multiscale stick-breaking mixture models.
\newblock {\em Statistics and Computing}, 31:13.

\bibitem[Tang et~al., 2022]{Tang2022}
Tang, T., Ghorbani, A., Squazzoni, F., and Chorus, C.~G. (2022).
\newblock Together alone: a group-based polarization measurement.
\newblock 56:3587--3619.

\bibitem[Tutz and Oelker, 2017]{Tutz_2017}
Tutz, G. and Oelker, M.-R. (2017).
\newblock Modelling clustered heterogeneity: Fixed effects, random effects and mixtures.
\newblock {\em International Statistical Review}, 85(2):204--227.

\bibitem[Ulker et~al., 2010]{Ulker_2010}
Ulker, Y., Günsel, B., and Cemgil, T. (2010).
\newblock Sequential monte carlo samplers for dirichlet process mixtures.
\newblock In Teh, Y.~W. and Titterington, M., editors, {\em Proceedings of the Thirteenth International Conference on Artificial Intelligence and Statistics}, volume~9 of {\em Proceedings of Machine Learning Research}, pages 876--883, Chia Laguna Resort, Sardinia, Italy. PMLR.

\bibitem[Uto, 2022]{Uto_2022}
Uto, M. (2022).
\newblock A bayesian many-facet rasch model with markov modeling for rater severity drift.
\newblock {\em Behavior Research Methods}.

\bibitem[Verbeke and Lesaffre, 1996]{Verbeke_1996}
Verbeke, G. and Lesaffre, E. (1996).
\newblock A linear mixed-effects model with heterogeneity in the random-effects population.
\newblock {\em Journal of the American Statistical Association}, 91(433):217--221.

\bibitem[Villarroel et~al., 2009]{Villarroel_2009}
Villarroel, L., Marshall, G., and Barón, A.~E. (2009).
\newblock Cluster analysis using multivariate mixed effects models.
\newblock {\em Statistics in Medicine}, 28(20):2552--2565.

\bibitem[Walker, 2007]{Walker_2007}
Walker, S.~G. (2007).
\newblock Sampling the dirichlet mixture model with slices.
\newblock {\em Communications in Statistics - Simulation and Computation}, 36(1):45--54.

\bibitem[Wang and Lin, 2014]{Wang_2014}
Wang, W.-L. and Lin, T.-I. (2014).
\newblock Multivariate t nonlinear mixed-effects models for multi-outcome longitudinal data with missing values.
\newblock {\em Statistics in Medicine}, 33(17):3029--3046.

\bibitem[Wirtz, 2020]{Wirtz2020}
Wirtz, M.~A. (2020).
\newblock {\em Interrater Reliability}, pages 2396--2399.
\newblock Springer International Publishing, Cham.

\bibitem[Zhang et~al., 2003]{Zang2003}
Zhang, C., Mapes, B.~E., and Soden, B.~J. (2003).
\newblock Part a no. 594 q.
\newblock {\em J. R. Meteorol. Soc}, 129:2847--2866.

\bibitem[Zhu et~al., 2021]{Zhu_2021}
Zhu, Y., Fung, A. S.-L., and Yang, L. (2021).
\newblock A methodologically improved study on raters’ personality and rating severity in writing assessment.
\newblock {\em SAGE Open}, 11(2):21582440211009476.

\bibitem[Zupanc and Štrumbelj, 2018]{Zupanc2018}
Zupanc, K. and Štrumbelj, E. (2018).
\newblock A bayesian hierarchical latent trait model for estimating rater bias and reliability in large-scale performance assessment.
\newblock {\em PLOS ONE}, 13(4):1--16.

\end{thebibliography}
\label{Appendix}
\section{Appendix}
 \label{sec:8} 
 \subsection{Remarks for multiple ratings}
When raters rate the same set of items $\mathcal{J}_i = \mathcal{J}$, $i=1,\dots,I$ a varying intercept can be identified for each item \citep{Patrícia_2023,Agresti2015,Nelson_2015}. These term might be added to equation \ref{HLM} (which is the same in both the standard and nonparametric formulation): \\
\begin{eqnarray}\label{HLM2}
    y_{ij}&=&\mathbf{x}_{ij}'\mathbf{\beta}+\mathbf{z}_{i}'\mathbf{u}_i+\delta_i+\epsilon_{ij},\;\; i = 1,..,I, \; j \in \mathcal{J}. \;\; 
    \end{eqnarray}
In both the standard HLM (i.e., assuming a multivariate normal distributed hierarchical rater effect) and the nonparametric HLM (i.e., specifying a DPM over the rater effect) the following distribution might be specified: 
 \begin{eqnarray}   
    \delta_j &\sim& N(0,\sigma_\delta^2) \quad  j= 1,..,J. \nonumber
\end{eqnarray}
where $\sigma_\delta >0$ is the scale parameter of $\delta$ and $J=|\mathcal{J}|$. See Section \ref{sec:2} and \ref{sec:3} for the other quantities and their distribution assumption. Specifying a conjugate prior for $\sigma_\delta$ additional steps might be added to the Gibbs sampling for the nonparametric HLM. \\
The main results of the present work and the interpretation of $\lambda$ (see Section \ref{sec:5}) still hold for this model specification. 

 \subsection{Details on Dirichlet Process Mixture}
 As noticed above, $\alpha$ is proportional to the \textit{concentration} of the realizations of $G$ in point masses. Indeed, considering the partition $(A,A^c)$ of $\Omega$, the variance of G(A) is defined as
\begin{center}
    $Var[G(A)]=\frac{G_0(A)(1-G_0(A))}{\alpha+1}$
\end{center}
\vspace{1.5px}
Thus, larger values of $\alpha$, conditioning on the number of raters $I$, reduce the variability of the DP, i.e. the process samples most of the time from $G_0$, $G$ tends to be an infinite number of point masses: the empirical distribution of $G$ tends to become a discrete approximation of the parametric $G_0$. In this case there is no a strong clustering since the probability of ties is very low. On the contrary, smaller values of $\alpha$ induce a strong clustering, the random weights distribution concentrate the probability mass to few points of the support of $G$ and the probability of ties is higher. Which in the present model means that several $u_i$ will be independent and identically distributed from a normal distribution indexed by the same parameters.
Moreover, Antoniak \citep{Antoniak_1974} demonstrated that 
\begin{center}
    $\mathbb{E}[C|I,\alpha]\displaystyle \approx \alpha$ ln$\left(\frac{I+\alpha}{\alpha}\right)$
\end{center}
\vspace{1.5px}
where $C$ is the number of clusters. Thus, the expected number of point masses of $G$ is proportional to both the $\alpha$ and the number of raters $I$. Every consideration regarding the role of the precision parameter on the distribution of $G$ should be conditioned to $I$. \\
 \subsection{Details on the Gibbs sampling}
 Further details regarding some parameters of the posterior sampling are showed as follow. \\
 The following matrix notation is here adopted: $\mathbf{X}_i=(\mathbf{x}_{i1}',\dots,\mathbf{x}_{i|\mathcal{J}_i|}')$, $\mathbf{Z}_i=(\mathbf{z}_{i1}',\dots,\mathbf{z}_{i|\mathcal{J}_i|}')$, are the design matrices for each rater $i=1,\dots,I$; and $\mathbf{X}=(\mathbf{X}_1,\dots,\mathbf{X}_I)$ and $\mathbf{Z}=diag(\mathbf{Z}_1,\dots,\mathbf{Z}_I)$ are the full design matrices.
\begin{enumerate}
    \item Referring to the non varying effects: \\
       \begin{eqnarray}
   \mathbf{b}_\beta^{*}&=& \left(\mathbf{B}_\beta^{-1} + \frac{1}{\sigma_\epsilon^2} \mathbf{X}'\mathbf{X}\right)^{-1} 
   \left(\mathbf{B}_\beta^{-1}\mathbf{b}_\beta + \frac{1}{\sigma_\epsilon^2} \mathbf{X}'\left(\mathbf{y}-\mathbf{Zu}\right)\right)
   \nonumber\\
   \mathbf{B}_\beta^{*}&=&  \left(\mathbf{B}_\beta^{-1} + \frac{1}{\sigma_\epsilon^2} \mathbf{X}'\mathbf{X}\right)^{-1}  \nonumber
\end{eqnarray} \\
    \item Referring to hierarchical effects: \\
\begin{itemize}
    \item For each rater $i =1,...,I$:
 \begin{eqnarray}
  \mathbf{\mu}_{c_i}^* &=& \left(\mathbf{D}_0^{-1} + \frac{1}{\sigma_\epsilon^2} \mathbf{Z}'_i\mathbf{Z}_i\right)^{-1} 
   \left(\mathbf{D}_0^{-1}\mathbf{\mu}_{c_i} + \frac{1}{\sigma_\epsilon^2} \mathbf{Z}'_i \left(\mathbf{y}_i-\mathbf{X}_i\mathbf{\beta}\right)\right)
   \nonumber\\
  \mathbf{Q}_{c_i}^* &=& \left(\mathbf{D}_0^{-1} + \frac{1}{\sigma_\epsilon^2} \mathbf{Z}'_i\mathbf{Z}_i \right)^{-1}  \nonumber
\end{eqnarray} 
\vspace{1.5px}
Here $\mathbf{\mu}_{c_i}$ is the location parameter vector of the cluster where the rater $i$ is allocated. \\
\item For each component $r = 1,...,R$ and each variable $d=1,...,q$, associated with an hierarchical effect: \\
 \begin{eqnarray}
 \mu_{0_r}^* &=& \left(\frac{c_r}{\sigma_{Q_d}^2} + \frac{1}{\sigma_{D_{0_d}}^2} \right)^{-1} 
 \left( \frac{c_r}{\sigma_{Q_d}^2} \overline{u}_{d,r} + \frac{\mu_{0_r}}{\sigma_{D_{0_d}}^2} \right)
 \nonumber\\
 \sigma_{D_{0_d}}^{2*} &=&  \left( \frac{c_r}{\sigma_{Q_d}^2} + \frac{1}{\sigma_{D_{0_d}}^2} \right)^{-1} \nonumber\\
  \sigma_{Q_{dr}}^2|\mathbf{\mu}, \mathbf{u} &\sim& IG \left( a_{Q_0} + \frac{r_c}{2},  b_{Q_0}+\frac{1}{2}\displaystyle\sum_{i=1}^{r_c} (u_{i_d}-\mu_{dr})^2 \right) \nonumber
\end{eqnarray} 
Here $\overline{u}_{dr}$ is the mean of the $d$-th hierarchical effect in the cluster $r$.\\
\item For rater $i =1,...,I$ and each component $r =1,...,R$:
 \begin{eqnarray}
 \mathbf{\omega}_{i_r}^*&=& \frac{\pi_r N_q(\mathbf{u}_i|\mathbf{\mu}_r, \mathbf{Q}_r)}{\sum_{r=1}^{R} \pi_r N_q(\mathbf{u}_i|\mathbf{\mu}_r, \mathbf{Q}_r)}. \nonumber 
\end{eqnarray} 
\end{itemize}
\end{enumerate}

\subsection{Some Trace Plots}
\begin{figure}
    \centering
    \includegraphics[scale=0.4]{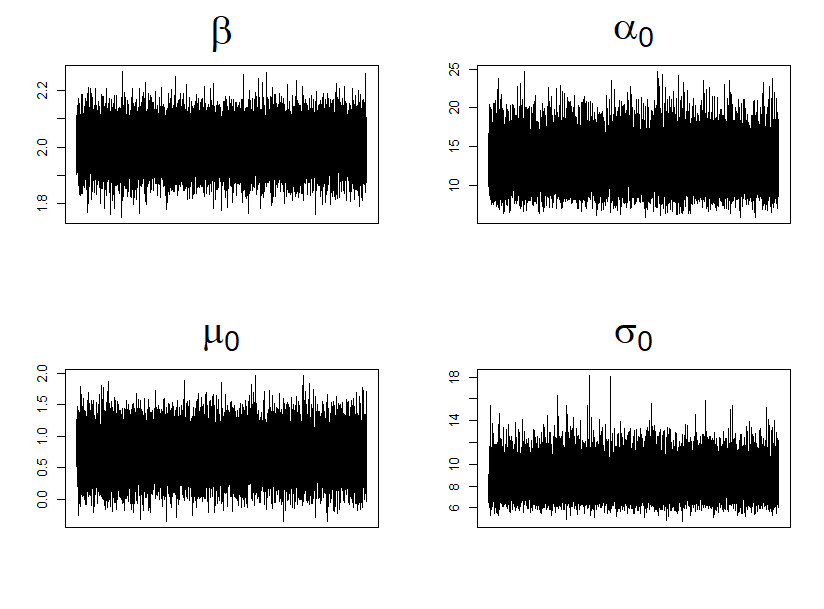}
    \caption{Trace Plots of some parameters from the second scenario. As it is shown they all converge properly.}
    \label{fig:3}
\end{figure}

\end{document}